%% file: main.tex
\documentclass{article}

\usepackage[utf8]{inputenc} 

\usepackage{amsfonts}       
\usepackage{nicefrac}       
\usepackage{microtype}      
\usepackage{lipsum}

\usepackage{color}
\usepackage{listings}
\usepackage{booktabs}
\usepackage{array}
\usepackage{graphicx}
\usepackage{longtable}

\usepackage{cite}
\usepackage{url}
\usepackage{fancyhdr}

\usepackage{soul}

\usepackage{float}
\usepackage{listings}
\usepackage{mdwmath}
\usepackage{mdwtab}
\usepackage{multirow}
\usepackage{multicol}
\usepackage{rotating}
\usepackage{setspace}
\usepackage[utf8]{inputenc}
\usepackage{lineno}
\usepackage{listings}

\usepackage{mdwmath}
\usepackage{mdwtab}
\usepackage{multirow}
\usepackage{multicol}
\usepackage{array}
\usepackage{booktabs}

\usepackage{enumitem}
\usepackage{xspace}
\usepackage[export]{adjustbox}
\usepackage{graphicx}
\usepackage{color,soul}
\usepackage{rotating}
\usepackage{setspace}
\usepackage{amsmath} 
\usepackage{amssymb}
\usepackage{float}
\usepackage{xcolor}

\usepackage{hyperref}
\usepackage{url}

\title{Patterns for Blockchain-Based Payment Applications}

\author{Qinghua Lu, Xiwei Xu, H.M.N. Dilum Bandara, Shiping Chen, Liming Zhu \\
Data61, CSIRO, Australia\\
University of New South Wales, Australia\\
firstname.lastname@data61.csiro.au}

\begin{document}

\maketitle

\begin{abstract}
As the killer application of blockchain technology, blockchain-based payments have attracted extensive attention ranging from hobbyists to corporates to regulatory bodies. Blockchain facilitates fast, secure, and cross-border payments without the need for intermediaries such as banks. Because blockchain technology is still emerging, systematically organised knowledge providing a holistic and comprehensive view on designing payment applications that use blockchain is yet to be established. 
If such knowledge could be established in the form of a set of blockchain-specific patterns, architects could use those patterns in designing a payment application that leverages blockchain. Therefore, in this paper, we first identify a token's lifecycle and then present 12 patterns that cover critical aspects in enabling the state transitions of a token in blockchain-based payment applications. 
The lifecycle and the annotated patterns provide a payment-focused systematic view of system interactions and a guide to effective use of the patterns. 

\end{abstract}

Blockchain, Pattern, Architecture, Payment, Token, Cryptocurrency, Escrow, Channel

\section{Introduction}
\label{intro}
Blockchain is an emerging distributed ledger technology that has attracted a broad range of interests from academia and industry to build the next generation of trustworthy applications. A large number of projects have been conducted to explore how to re-architect systems, and to build new applications and business models using blockchain as a general, decentralised computing and storage environment through the advent of smart contracts (i.e., programs on a blockchain)~\cite{2019-Bratanova-ACS}.

Payment is considered the killer application of blockchain technology, which executes monetary transactions for goods or services previously agreed upon by all parties involved. The ``money'' used on a blockchain includes both cryptocurrencies and tokens. \emph{Cryptocurrencies} are baked into the blockchain platforms as the base currency (e.g., Bitcoin and Ether) for transactions and provide incentives for blockchain platform operations (e.g., mining). \emph{Tokens} are application-tier coins that are created and exchanged using smart contracts on the blockchain. Tokens may represent cryptocurrencies (e.g., Binance Coin and Uniswap) or used as a proxy for fiat currency (e.g., Tether and US Dollar Coin).



\begin{figure}[t]
	\centering
	\includegraphics[width=0.4\columnwidth]{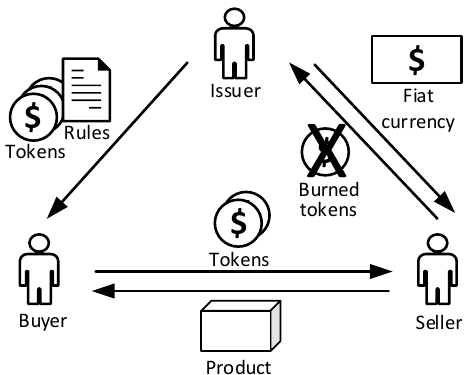}
	\caption{Overview of token-based payments.}
	\label{scenario}
\end{figure}

Fig.~\ref{scenario} illustrates an overview of token-based payments. A token \emph{issuer} issues a new token associated with a set of spending rules. One or more tokens are allocated to a \emph{buyer}. For example, in food stamps, the issuer is the government agency that manages the program and buyers are the people classified as food insecure. The spending rules may specify eligible kinds of food, their quantities, and sellers. A buyer then purchases food products from a \emph{seller} using the allocated tokens. A seller could be a supermarket that cashes out multiple tokens from the government agency. However, as different food stamps may carry different spending rules, it is in the seller's best interest to validate them at the time of payment. This ensures that the seller can convert the collected tokens to fiat currency for the products it sold. 
Alternatively, the buyer may also want to know the authenticity of the product and whether the seller has the authority to sell it. 
Once redeemed, the issuer needs to destroy the tokens (aka., burn token) to ensure the tokens are not reused (i.e., double spend). 

As discussed in the above example, payments are usually conditional, where they are made subject to a set of conditions. Payment conditions could be complicated in many environments, and could range from coupons with a designated product/service and an expiration date to insurance payouts and social welfare benefits. Further, the conditions need to be specified in advance (usually with the consent of parties involved in the transaction) such that they can be checked at the time of payment. Thus, in such use cases, the currency needs to be programmable to specify payment conditions, enable automatic validation and enforcement, and support immutable record-keeping of associated transactions. These properties are well aligned with the design of tokens realised by blockchains (i.e., provides immutable record-keeping and multi-party transactions) and smart contracts (i.e., provides programmability and automatic enforcement). 


As blockchain technology is still emerging, systematically organised knowledge providing a holistic and comprehensive view on designing blockchain-based payment applications is yet to be established. Therefore, to form such knowledge, in this paper, based on a literature review and our project experience, we collect a set of 
patterns that cover the critical aspects in the design of blockchain-based payment applications. We identify the lifecycle of a payable token, in which the identified patterns are associated with the state transitions along the lifecycle. The lifecycle and the annotated patterns provide a payment-focused systematic view of system interactions and a guide to the effective usage of the patterns to both architects and developers. In this paper, we focus on blockchain-based payable tokens such as those based on  ERC-1363\footnote{\url{https://eips.ethereum.org/EIPS/eip-1363}} standard that can be spent to purchase products or services. We also consider native cryptocurrencies, fungible tokens (e.g., ERC-20\footnote{\label{erc_20}\url{https://eips.ethereum.org/EIPS/eip-20}}), and non-fungible tokens (e.g., ERC-721\footnote{\label{erc_721}\url{https://eips.ethereum.org/EIPS/eip-721}}) that are used within smart contracts to attach to a specific buyer (e.g., a voucher given to a patient), specify payment conditions, and execute payments.

The remainder of this paper is organised as follows: Section~\ref{related} discusses related work. The overview of payment patterns is introduced in Section~\ref{patternoverview}. Section~\ref{patterns} presents each pattern in details. Section~\ref{conclusion} gives concluding remarks.

\section{Related Work}
\label{related}
Patterns have been collected to document the reusable solutions for designing blockchain-based applications~{\cite{xu2018pattern, eberhardt2017or, smartContractICBC, wohrer2018smart, SSIpattern, data_migration, zhang2017applying, oracle}}. 
Some of the blockchain patterns focus on the design for generic applications~\cite{xu2018pattern, oracle, eberhardt2017or}, while others are collected only for specific applications~\cite{SSIpattern, data_migration, zhang2017applying}. 
Eberhardt and Tai \cite{eberhardt2017or} identified five data and computation design patterns for blockchain-based applications and discussed what data and computation to be kept on-chain and what to be placed off-chain.

Many patterns are summarised for smart contracts~\cite{smartContractICBC, wohrer2018smart, bartoletti2017empirical, Wohrer_Zdun}. For example, Liu et al.~\cite{smartContractICBC} presented eight design patterns for writing smart contracts. Wöhrer and Zdun \cite{wohrer2018smart} summarised six design patterns to improve the security of smart contracts. 
Further, empirical studies on smart contracts have been performed to identify the common smart contract design patterns~\cite{bartoletti2017empirical, Wohrer_Zdun}. 
There are also examples of applications recognising the adoption of specific pattern collections to showcase their robust designs. For example, OriginChain \cite{IEEESoftware2017, originChain} is a supply chain application that adopts data management and smart contract design patterns to improve system adaptability. uBaaS is a service platform that encapsulates patterns as services to generate code skeletons to facilitate the development of blockchain-based applications \cite{ubaas}. In \cite{decision_model}, we proposed a decision model that assists developers and architects in selecting a set of patterns for blockchain-based applications. 

There have been significant efforts in both industry and academia to address payment challenges using the blockchain technology. In 2018, CSIRO’s Data61 and Commonwealth Bank of Australia developed a proof of concept app to create and manage programmable ``smart money'' using blockchain technology~\cite{smartmoney} through the case study of the National Disability Insurance Scheme (NDIS). In 2019, TietoEVRY and Kela (The Social Insurance Institution of Finland) had cooperated to develop a proof of concept prototype for smart money~\cite{smartmoneyfinland}. Goldfeder et al.~\cite{escrow} studied the escrow problem for physical products and design a set of escrow schemes for Bitcoin.  

Compared to the existing literature and application of known patterns, our study focuses on the patterns in the context of blockchain-based payments, which have not been covered before. Although patterns such as \emph{burned token}, \emph{escrow}, and \emph{oracles} are already discussed in previous work~\cite{xu2018pattern, SSIpattern}, in this paper, we adapt them in the area of payments and aim to provide a holistic and comprehensive guide on the design of blockchain-based payment applications.

\section{Overview of Blockchain-based Payment Patterns}
\label{patternoverview}

A variety of constraints need to be satisfied while building payment applications on a blockchain. While each application is unique, we can always learn from the commonly recurring problems and the corresponding reusable design solutions. Those reusable design solutions can be established as a set of design patterns.

The payment patterns are derived from our blockchain project experience and literature review.
We started with identifying our research question, which is ``What are the effective patterns for designing a blockchain-based payment application?''. We used Google Search and Google Scholar as the source databases. The search keywords include ``blockchain'', ``payment'', ``smart contract'', ``token'', ``asset'', and their combinations. We also used snowballing to expand the pool of source literature. Next, we filtered the sources according to the predefined selection criteria, e.g., ``payment must include ownership transfer of money''. Finally, we used open and axial coding to analyse the selected sources and derive related patterns iteratively.

\begin{table*}[tbp]
\footnotesize
\centering
\caption{Overview of blockchain payment patterns.}
\label{overview}
\begin{tabular}{p{0.09\columnwidth}p{0.125\columnwidth}p{0.625\columnwidth}}
\toprule

\multicolumn{1}{l}{\bf Category} &
\multicolumn{1}{c}{\bf Name} &
\multicolumn{1}{c}{\bf Summary}
\\
\midrule

\multirow{12}{0.09\columnwidth}{\textbf{Token\\Design Patterns}} & Token Template & Tokens are designed in a customised way using a \textit{token template} that can be extended or instantiated.
\\
\cmidrule(l){2-3}
& Token Registry & A token registry maintains the ownership information of tokens through a mapping of token ID and wallet address.
\\
\cmidrule(l){2-3}
& Policy Contract & A \textit{policy contract} specifies the policies that a token must satisfy, e.g., whether a token must be spent on the product or services supplied by eligible sellers or how many tokens can be spent in a transaction.
\\

\cmidrule(l){2-3}

& Burned Token & When redeemed, a spent token is marked as unusable/unspendable (aka., \textit{burned}).
\\



\cmidrule(l){1-3}
\multirow{27}{0.09\columnwidth}{\textbf{Payment Management Patterns}} & \multirow{4}{0.125\columnwidth}{Escrow} & Before making a transaction, tokens are transferred to a third-party smart contract called the escrow. The escrow holds the deposited tokens until the payment conditions are satisfied.
\\

\cmidrule(l){2-3}
& \multirow{4}{0.125\columnwidth}{Payment Channel} & 
A two-way pathway (aka., \textit{payment channel}) is established between transacting parties to perform transactions off-chain while recording opening and final settlement transactions on the blockchain. 
\\

\cmidrule(l){2-3} 
& Seller Credential & The buyer can validate the seller's business qualification through seller credentials. 
\\
\cmidrule(l){2-3} 
&\multirow{3}{0.125\columnwidth}{Stealth Address} &  A \textit{stealth address} protects the privacy of the parties involved in the payment by creating a one-time address for the transaction. 
\\
\cmidrule(l){2-3}
& \multirow{3}{0.125\columnwidth}{Oracle} & To examine the fulfilment of payment conditions, an \textit{oracle} is used to introduce the state of external systems 
into the closed blockchain execution environment. 
\\
\cmidrule(l){2-3}
& \multirow{3}{0.125\columnwidth}{Multi-signature} & Given a pool of $n$ addresses that could authorise a payment, a subset $m$ of them can authorise a payment by signing the respective transaction ($m \leq n$). 
\\
\cmidrule(l){2-3}
& Token Swap & Token swap allows users to trade directly between two types of tokens as an atomic transaction. 
\\
\cmidrule(l){2-3}
& Authorised Spender & An authorised third-party spender is allowed to spend a certain amount of tokens on behalf of the approver. 
\\

\bottomrule
\end{tabular}
\end{table*}

\begin{figure*}[!ht]
	\centering
	\includegraphics[width=\textwidth]{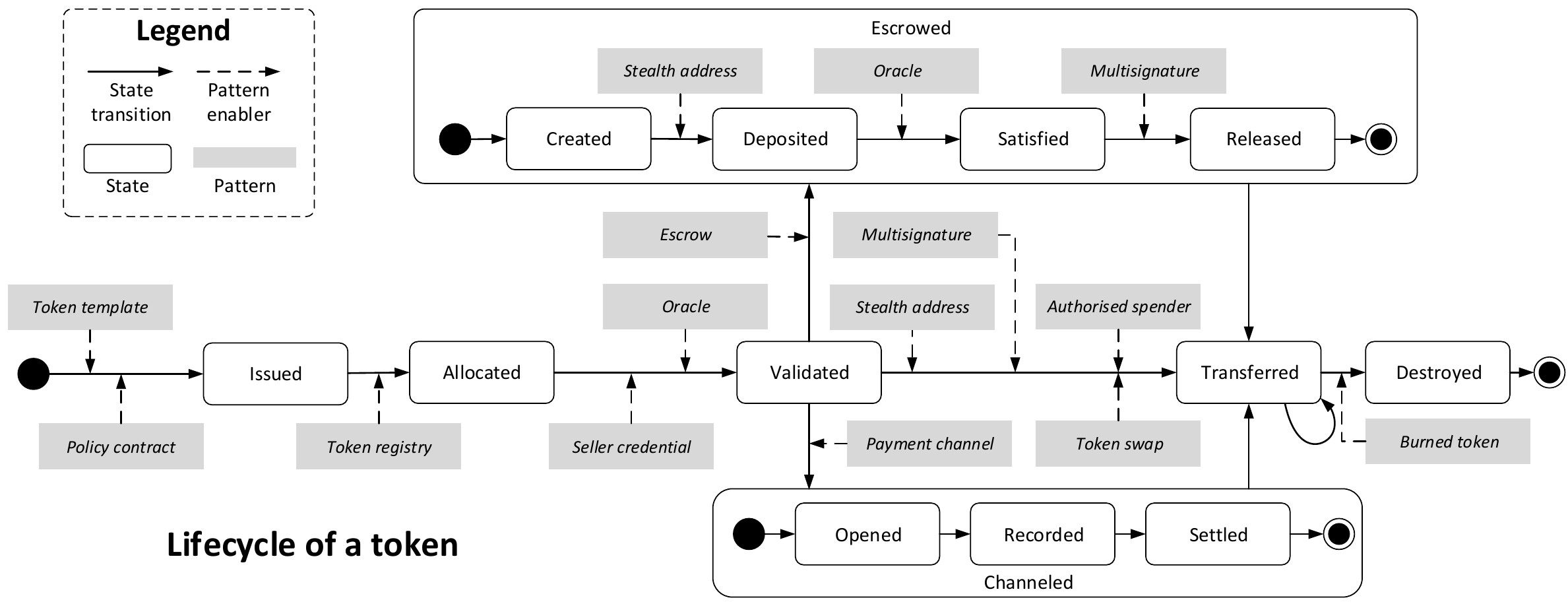}
	\caption{Lifecycle of a token with the annotated patterns.}
	\label{lifecycle}
\end{figure*}

Table~\ref{overview} presents an overview of the collected patterns. We classify the patterns into two categories: token design and payment management patterns. We provide a summary for each pattern. 
Although some patterns (e.g., \textit{burned token}, \textit{stealth address}, and \textit{multisignature}) are provided as embedded features by some blockchain platforms, they can still be considered at the application-level as architects and developers can implement those features in smart contracts and applications. For example, the burning of native tokens like BNB is baked into blockchain platform logic, while ERC-20 like tokens could be burned by the smart contracts that created them. Hence, such tokens are at the control of the application developer. Similarly, while token swap could be supported at the blockchain platform level, it becomes more complicated when token are swapped across blockchains. In such cases, an application-level solution, such as Hash-Lock Time-Lock Contracts (HTLCs), is needed to link two blockchains. Therefore, while we focus on patterns that assist the blockchain-based payment application design in this paper, some of those patterns may also be applied to a different level.

Fig. \ref{lifecycle} illustrates the lifecycle of a token and highlights the patterns associated with state transitions within the lifecycle. The lifecycle starts with the issuance of a token. The token can be designed in a customised way using a \textit{token template} pattern which specifies necessary details (e.g., token name, symbol, number of decimal places, and the total number of tokens supplied) and advanced features (e.g., the minting process, initial distribution, and token burning process). 
A \textit{policy contract} pattern stipulates the rules for using a group of tokens such as how many vouchers can be spent in a transaction and whether the token must be spent on the specified products or services supplied by eligible sellers. 

Once issued, the predefined number of tokens are allocated to the buyer's wallet addresses. 
The \textit{token registry} pattern could be used to maintain the token ownership information as a mapping between a wallet address and the number of tokens or token identifiers associate with it. The buyer's wallet receives the allocated tokens and sends a notification to the consumer that the tokens are received. 

Before purchasing a product or service, a consumer's wallet could use the \textit{seller credential} pattern to validate the seller's business qualifications. 
Further, the \textit{oracle} pattern can introduce the external state/data into the blockchain execution environment, which is self-contained. For example, a payment may be subject to an external state confirmed by a human like a truck driver confirming the collection of an item or an IoT device automatically reporting quality attribute of the item or its ambient conditions through the oracle. 

Once the required transaction information is validated, the token(s) are transferred to the seller. The \textit{escrow} pattern can be used for conditional payments, which is suitable for high-value, infrequent payments that require strong security guarantees. The escrow holds the deposited tokens until the payment conditions are satisfied. 
\textit{Multisignature} pattern can be applied to sign transactions requiring the approval of multiple parties, e.g., to control the release of tokens held by an escrow. Please note that \textit{stealth address}, \textit{oracle}, and \textit{multisignature} patterns can be applied to a transaction with or without an escrow.

To deal with frequent, low latency, and low-value payments, a \textit{payment channel} can be designed to perform transactions off-chain with opening and final settlement transactions recorded on-chain. The \textit{authorised spender} pattern is used to allow a third party (e.g., a smart contract) to spend a certain amount of tokens on behalf of the approver. \textit{Token swap} pattern enables users to trade directly between two types of tokens as an atomic swap. The \textit{stealth address} pattern enables a given payment to be made to a one-time account/address, such that the anonymity set is only provided to the parties involved in the transaction. Once the transaction is settled, the seller is notified that the payment has been executed successfully.


Depending on the use case, tokens could be transferred multiple times. Tokens might need to be redeemed in some use cases. Thus, the spent/redeemed tokens might become \textit{burned tokens}, which means the tokens are no longer usable. 

\section{Blockchain-Based Payment Patterns}
\label{patterns}
In this section, we present patterns that shape the architectural elements and their interactions in blockchain-based payment applications. Each pattern is described following the extended pattern form in \cite{meszaros1998pattern}, which includes the pattern name, summary, problem statement, forces that make the problem difficult to solve, solution, consequences of applying the solution, and real-world known uses. The proposed patterns' intended audience is software architects and developers who have to deal with blockchain-based payment application design and development.

\subsection{Token Template}
\textbf{Summary:}
Tokens can be designed in a customised manner using a template smart contract that can be extended or instantiated.

\vspace{0.5em}\noindent \textbf{Context:}
Blockchain provides a trustworthy platform to realise tokenisation. In a blockchain application, a token represents a programmable digital unit of value (i.e., assets) recorded on the blockchain for a specific purpose, like buying a product or consuming a service. The same application may use different tokens or asset classes, e.g., a food stamp issued for a child is usually different from an adult.

\vspace{0.5em}\noindent \textbf{Problem:}
A token is usually defined as a data structure embedded in a smart contract to represent an asset. How can developers design and develop different types of tokens with various features? 

\vspace{0.5em}\noindent \textbf{Forces:}
\begin{itemize}[leftmargin=*]
    \item \textit{Flexibility.} A blockchain application might support different types of tokens with different properties and rules.
    \item \textit{Interoperability and liquidity.} Tokens supported by different applications might need to be exchanged frequently.    
    \item \textit{Vulnerabilities.} Handling tokens with a high value is risky, as lost tokens cannot be replaced/recovered on the blockchain. Further, vulnerabilities or bugs might be introduced when developing a blockchain-based application. 
    As blockchain is an immutable data store, updating already deployed smart contracts is impossible. This makes it difficult to fix bugs by releasing new versions of a smart contract. 
    \item \textit{Productivity.} Blockchain is an emerging technology with limited tooling and documentation. Thus, developers can have a steep learning curve, which affects the developer productivity.
    \item \textit{Cost.} On a public blockchain, transaction fees need to be paid to deploy and execute smart contracts. The execution fee is proportional to the computational and storage complexity of the smart contract.
\end{itemize}


\vspace{0.5em}\noindent \textbf{Solution:}
Blockchain-based payment applications might need to issue different types of tokens with customisation. However, as most tokens have a common set of properties and functions, a token can be defined as a template smart contract. The template can then be either extended or instantiated by another smart contract to issue a new token at run time. Multiple tokens may also be issued based on the same template by setting different properties.
There can be two types of tokens: cash tokens and voucher tokens. \textit{Cash tokens} are fungible tokens that are fully exchangeable with each other and can be divided into smaller denominations for multiple transactions. In contrast, \textit{voucher tokens} are non-fungible tokens that are unique and must be spent in a single transaction regardless of the price of a product or service. Both cash tokens and voucher tokens can be developed as token templates. 

\begin{figure}[t]
	\centering
	\includegraphics[width=0.8\columnwidth]{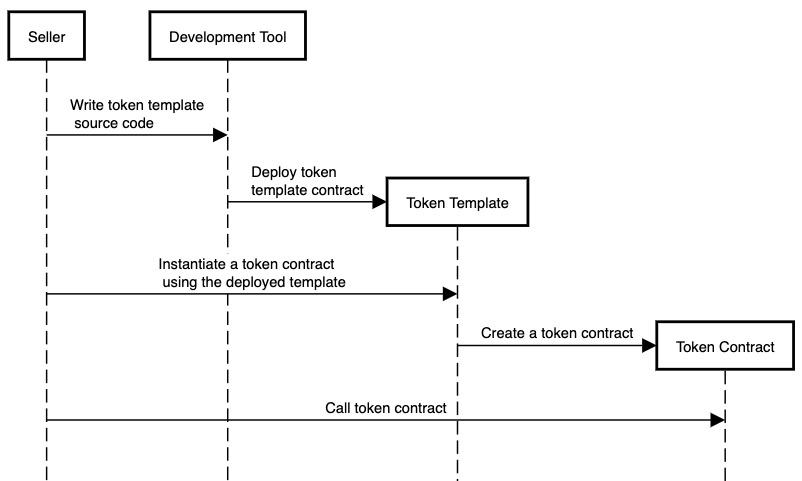}
	\caption{Sequence diagram of token template.}
	\label{template_seq}
\end{figure}

Fig. {\ref{template_seq}} shows the sequence of activities required to create and utilise a token template. In the token template smart contract, the user can define the token's attributes and methods to manage the token. 
The common attributes usually consist of essential information such as token name, symbol, number of decimal places, and type. Other attributes such as the minting process, initial distribution of tokens, and burning process could be specified depending on the token's intended use. The specified methods enforce rules that govern how the tokens are minted, owned, and transferred. Optional attributes and methods in the template allow customisation. If that is insufficient, the template smart contract can be extended by another smart contract to support customised attributes and methods. Once developed, the template would change rarely. Thus, it can be thoroughly tested and certified to be free of defects before the deployment. 
Once developed, deploy the template to the blockchain as a factory smart contract. The template contract can be made a factory contract by adding a public method to spawn new child contracts with given properties. Whenever a new token type is to be created, issue a transaction with relevant properties to the spawn function in the factory contract. In turn, the factory contract instantiates the template with the given properties to spawn the token contract. Subsequently, transactions could be issued to the token contract to transact using the newly created tokens. Alternatively, if the new token contract is developed by extending the template, the template can be kept off-chain. The new token contract can then be directly deployed on to the blockchain, which internally carries template contract code (i.e., parent contract).

\vspace{0.5em}\noindent \textbf{Consequences:}

\vspace{0.5em}\noindent Benefits:
\begin{itemize}[leftmargin=*]
\item \textit{Productivity and security.} A token template can support developing a particular type of token by creating an instance of the token template. A well-defined and tested template simplifies blockchain application development, reduces time, and enhances security and reliability.
\item \textit{Flexibility.} A blockchain application might support different types of tokens with different rules. The use of token templates in the design can improve modularity.
\item \textit{Interoperability and liquidity.} As the token template can be designed based on a token standard, 
it offers high interoperability and liquidity through token transfer and swap.

\end{itemize}

\vspace{0.5em}\noindent Drawbacks:
\begin{itemize}[leftmargin=*]
\item \textit{Cost.} If a public blockchain is used, an extra cost is required to deploy the token template contract on the blockchain and call its token contract instance creation function. Further, the generalisation of the template increases the cost of computational and storage complexity of the smart contract, increasing the transaction fees needed to deploy and manipulate it.
\item \textit{Productivity.} It takes time to design, develop, and test a token template carefully. However, this is a one time process.

\end{itemize}

\vspace{0.5em}\noindent \textbf{Related patterns:} 
The \textit{token registry} pattern can be used to maintain the ownership information of tokens generated using the token contract. Terms and conditions that a generated token needs to satisfy can be specified by linking the token to policies specified using the \textit{policy contract} pattern. Once redeemed, a spent token can be destroyed or marked as unusable using the \textit{burned token} pattern.

\vspace{0.5em}\noindent \textbf{Known uses:}
\begin{itemize}[leftmargin=*]
    \item \textit{ERC-20}\footnote{See footnote \ref{erc_20}} and \textit{ERC-721}\footnote{See footnote \ref{erc_721}} have emerged as the de-facto standards for implementing fungible and non-fungible tokens in Ethereum and sister blockchains, respectively, by instantiating reference implementations like  OpenZeppeline\footnote{\url{https://docs.openzeppelin.com/contracts}}.
    \item \textit{Lorikeet}~\cite{lorikeet2020} is a model-driven engineering tool that provides off-chain templates for the developers to customise fungible and non-fungible asset data contracts based on ERC-20 and ERC-721 interface definitions.
    \item \textit{Token Mint}\footnote{\url{https://tokenmint.io/}} is a tool to create 
    fixed supply and mintable ERC-20 and ERC-777 tokens and crowd sale contracts without coding.
    \item \textit{Token Launcher}\footnote{\url{https://thetokenlauncher.com/}} creates mintable ERC-20 tokens by setting the desired specification.

\end{itemize}

\subsection{Token Registry}
\textbf{Summary:} A token registry maintains the ownership information of tokens by mapping the token identifier and wallet address.

\vspace{0.5em}\noindent \textbf{Context:} A buyer (i.e., token owner) needs to pay a certain number of tokens to purchase a product or service. When a payment is made, the respective tokens' ownership needs to be transferred from the buyer to the seller. Further, tracking historical ownership of tokens is required to ensure traceability. The payment application needs to manage a large number of transacting parties, tokens, and transactions.

\vspace{0.5em}\noindent \textbf{Problem:}
How can a large number of tokens, their owners, and transactions be effectively tracked?

\vspace{0.5em}\noindent \textbf{Forces:}
\begin{itemize}[leftmargin=*]
    \item \textit{Dynamism.} The ownership of tokens could change frequently.
    \item \textit{Scalability.} A large number of token and their owners need to be mapped.
    \item \textit{Transparency.} Transfer of token ownership needs to be tracked for compliance auditing.
\end{itemize}


\begin{figure}[t]
	\centering
	\includegraphics[width=0.7\columnwidth]{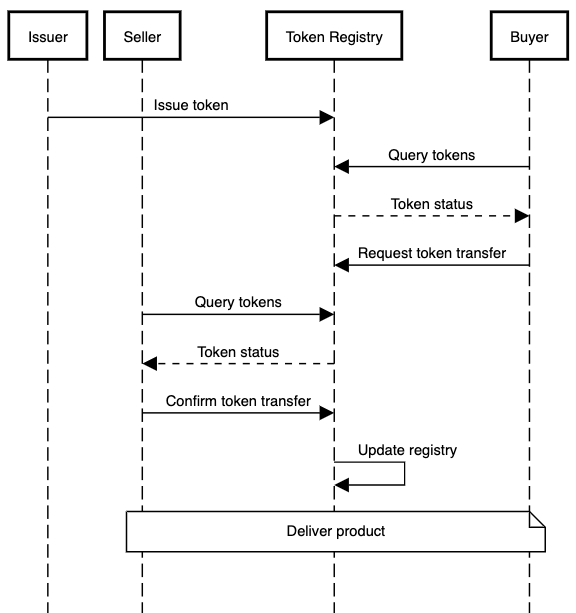}
	\caption{Sequence diagram of token registry.}
	\label{token_seq}
\end{figure}

\vspace{0.5em}\noindent \textbf{Solution:} 
A table in the form of a token registry contract can be used to record the ownership information of tokens by mapping token identifier (ID) to the owner's wallet address. 
As seen in Fig. {\ref{token_seq}, the issuer first registers both the token ID and wallet address of the buyer as variables in the token registry smart contract. 
The registry may also keep track of the token balance. This registry provides the ground truth on token ownership, as it could be manipulated only through the smart contract by authorised parties. For example, while both the buyer and seller can verify their token status, only the buyer can initiate a request to transfer a token. When a product to be purchased, the buyer requests the registry to transfer the desired number of tokens. As it is in the seller's best interest to ensure it receives only the accepted types of tokens, the seller could then confirm the token transfer to indicate its willingness to accept the tokens. Once confirmed, the registry smart contract updates the buyer's and seller's token balances. Delivery of product is handled off-chain.}

Writing permission to the registry can be added through a permission control module. For example, only the issuer should register tokens and perform the forced transfer for legal action or fund recovery. Further, it can be specified that ownership transfer should proceed only if the token is in the spendable state or once the seller approves. Such conditions can be checked through a method in the registry. If the conditions are not met, the registry can throw an error rather than receiving the token. This pattern can be generalised to keep track of eligible sellers and a matrix that maps products and sellers. For example, a registry contract can maintain the seller attributes such as names, wallet addresses, certifications, and accolades.

\vspace{0.5em}\noindent Benefits:
\begin{itemize}[leftmargin=*]
\item \textit{Traceability.} Traceability is enhanced, e.g., how many tokens are owned by whom at what time is reflected in the registry.
\item \textit{Upgradability.} Because the token ownership and balance data are managed separately by the token registry, the rest of the smart contract code can be upgraded without changing the token registry contract. 
\item \textit{Cost.} By separating token ownership data from the application code, there is no additional cost for data migration when the application logic is upgraded.
\end{itemize}

\vspace{0.5em}\noindent Drawbacks:
\begin{itemize}[leftmargin=*]
\item \textit{Cost.} Recording ownership data in smart contract incur an additional cost if a public blockchain is adopted. Further, there could be additional cost for
updating the registry while executing a transaction.
\end{itemize}

\vspace{0.5em}\noindent \textbf{Related patterns:} 
The \textit{token template} pattern can use the token registry to maintain the ownership information of tokens generated using the token contract. It can also be used to build the credentials registry of sellers used in the \textit{seller credential} pattern. Once redeemed, a spent token can be destroyed or marked as unusable by the registry using the \textit{burned token} pattern.

\vspace{0.5em}\noindent \textbf{Known uses:}
\begin{itemize}[leftmargin=*]
    \item \textit{Perth Mint Gold} token~\cite{Perth_Mint} uses a registry contract to map physical gold ownership (specified via GoldPass certificates) to the owner's Ethereum address.
    \item \textit{Lorikeet}~\cite{lorikeet2020} is a model-driven engineering tool that generates centralised and distributed asset registry smart contracts compliant with ERC-20 and ERC-721 standards. It could further specify relationships and many-to-many mappings between attributes (within the same or a different registry), enforce access control, and track the history of changes made to assets tracked by the registry.
    \item \textit{Parity Token Registry}\footnote{\url{https://github.com/openethereum/token-registry}} enables all the registered token to be visible to Parity wallet users.
    \item \textit{Codefi}\footnote{\url{https://github.com/ConsenSys/UniversalToken}} and \textit{Cardano}\footnote{\url{https://github.com/cardano-foundation/cardano-token-registry}} provide an ownership registry for Universal and Ada tokens, respectively.
    \item \textit{Making Money Smart}\footnote{\label{making_money_smart}\url{https://commbank.com.au/business/business-insights/making-money-smart.html}}, \textit{Stadjerspas Smart Vouchers}\footnote{\url{https://stadjerspas.nl/}}, and \textit{Pension Infrastructure}\footnote{\url{https://apg.nl/en/publication/apg-and-pggm-develop-a-blockchain-application-for-pension-administration/}} use registries to specify eligible sellers and fund providers.
\end{itemize}

\subsection{Policy Contract}
\textbf{Summary:}
A \textit{policy contract} specifies the policies that a token must satisfy, e.g., whether a token must be spent on the product or services supplied by eligible sellers or how many tokens can be spent in a transaction. 

\vspace{0.5em}\noindent \textbf{Context:}
A payment should be settled only when specific business rules or policies are satisfied. For example, government support programs or charitable donations usually indicate limits and constraints regarding products/services that could be purchased, their quantities, and eligible sellers.

\vspace{0.5em}\noindent \textbf{Problem:}
How to specify policies governing the use of one or more tokens and validate them by the payment applications before allowing a payment?

\vspace{0.5em}\noindent \textbf{Forces:}
\begin{itemize}[leftmargin=*]
    \item \textit{Diversity.} A variety of policies often govern the use of a token. They are challenging to find and interpret in relation to one another.
    \item \textit{Dynamism.} The policies or business rules are dynamically attached/removed to/from tokens. 
    \item \textit{Complexity.} The policies or business rules are complex and difficult to codify. 
\end{itemize}


\begin{figure}[t]
	\centering
	\includegraphics[width=0.7\columnwidth]{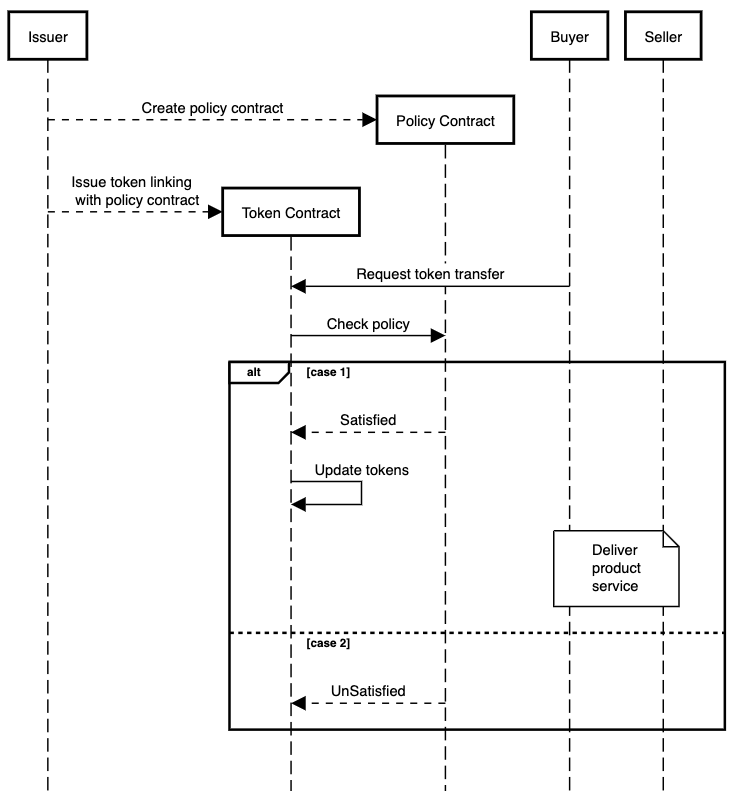}
	\caption{Sequence diagram of policy contract.}
	\label{policy_seq}
\end{figure}

\vspace{0.5em}\noindent \textbf{Solution:} A policy smart contract can be designed to specify policies that a token must satisfy before use. Fig. {\ref{policy_seq}} shows
the sequence of activities to specify and use a policy contract. First, the issuer creates a policy smart contract and deploys it to the blockchain. For example, the contract could specify that an attached token is spendable only on medicine, and the corresponding policy check logic could validate whether a given pharmacy is registered in the approved seller registry (i.e., a registry smart contract that maintains the information of eligible sellers from whom the token can be spent to purchase that medicine). Then the token contract can link the deployed policy to any newly issued token. When the buyer requests the token transfer during payment, the token contract (or token registry, not shown in the figure) can validate the policy by calling the policy contract. When all the policy conditions are satisfied, the token balance is updated on the token contract (or token registry). If the payment conditions are unsatisfied,   the transaction is rejected.

The policy check logic could be implemented as smart contract methods that are invoked before the payment function. It is also possible to specify a generic method to validate a policy document given in a pre-specified format. For example, JSON schema could be used to specify policies that can be codified as key-value pairs.
The mapping of policy contracts can happen at the level of either a single or group of tokens. In some instances, policies are expected to remain the same until the token is redeemed (or at least within a given time window). In other cases, the policies governing the tokens may change with time (e.g., changes in service agreements). Thus, the issuers may want to enforce the latest policy at the time of spending. Typically, a policy contract needs to be designed with a generic interface such that it can be dynamically attached/detached to/from the token. The attachment can be specified as a tuple consisting of the address of policy smart contract and list of policy functions.

\vspace{0.5em}\noindent Benefits:
\begin{itemize}[leftmargin=*]
\item \textit{Dynamism.} Policy contracts can be flexibly attached, updated, or removed from a token.
\item \textit{Diversity.} Policy contracts enable issuers to specify a variety of policies in policy contracts for the usage of different tokens. 
\item \textit{User Productivity.} Multiple policies and tokens could be automatically checked using smart contracts without needing the buyer and seller to interpret them manually. 
\end{itemize}

\vspace{0.5em}\noindent Drawbacks:
\begin{itemize}[leftmargin=*]
\item \textit{Cost.} If a public blockchain is used, an extra cost is required to deploy a policy contract on blockchain and call its functions at the time of transacting.
\item \textit{Developer Productivity.} Policies are challenging to codify; hence, it takes time to carefully design, develop, and test a policy contract. However, this is a one time process.
\end{itemize}

\vspace{0.5em}\noindent \textbf{Related patterns:} \textit{Token template} and \textit{token registry} patterns can use the policy contract to link and validate the policies that a token needs to satisfy. A policy contract can also be referred when a third party conducts a transaction using the \textit{authorised spender} pattern.

\vspace{0.5em}\noindent \textbf{Known uses:}
\begin{itemize}[leftmargin=*]
    \item \textit{Making Money Smart}\footnote{See footnote \ref{making_money_smart}}\cite{smartmoney} is a joint pilot project by the Commonwealth Bank of Australia and CSIRO's Data61 
    that attaches policy contracts to each pouch of tokens issued to redeem National Disability Insurance Scheme (NDIS) benefits in Australia. The policy defines what healthcare services could be used, their quantities, and eligible sellers. 
    \item \textit{Smart Money}\footnote{\label{tietoevry}\url{https://tietoevry.com/en/campaigns/2020/smartmoney--a-conditional-digital-payment-guarantee/}}\cite{smartmoneyfinland} is a blockchain-based payment platform jointly developed by Kela (The Social Insurance Institution of Finland) and TietoEVRY in cooperation with the Finnish Tax Administration, Financial Supervisory Authority of Finland, and Borenius Attorneys. Spending rules are defined as constraints for smart money tokens and are verified at the time of a transaction. 
    \item \textit{ERC-1400}\footnote{\label{erc_1400}\url{https://github.com/ethereum/EIPs/issues/1411}} is a library of standards for security tokens on Ethereum that specify transfer restrictions, rights, and obligations. Polymesh\footnote{\label{polymesh}\url{https://developers.polymesh.live/originate/sdk}} specify how those restrictions can be specified using JSON schema in a KYC (Know Your Customer) setting.
    \item \textit{CAIPY}~\cite{carpolicy2018} is a blockchain-based car insurance policy framework. In the design of \textit{CAIPY}, policy contracts model process fragments defined by the physical insurance policies between customers and insurers. The policy contracts in \textit{CAIPY} only govern the conditions for claims, not the use of tokens as insurance payments.
\end{itemize}

\subsection{Burned Token}
\textbf{Summary: } Once the redeem process is finalised, the spent tokens become unspendable. 

\vspace{0.5em}\noindent \textbf{Context:}
Once the sellers receive tokens from the buyers, they need to redeem the tokens for fiat currency (e.g., US Dollars). Once the tokens are redeemed in return for a fiat currency payment, they should not be further usable (i.e., prevent double-spending).

\vspace{0.5em}\noindent \textbf{Problem:}
How to ensure that a token is unspendable once it is redeemed?

\vspace{0.5em}\noindent \textbf{Forces:}
\begin{itemize}[leftmargin=*]
    \item \textit{Double spending.} A token should be redeemed only once.
    \item \textit{Complexity.} There are a large number of redeemed tokens that need to be maintained and tracked. 
    \item \textit{Vulnerability.} Vulnerability might exist if the token contract or the token registry is not appropriately managed or contains bugs, e.g., the private key is stolen. 
\end{itemize}


\begin{figure}[t]
	\centering
	\includegraphics[width=0.7\columnwidth]{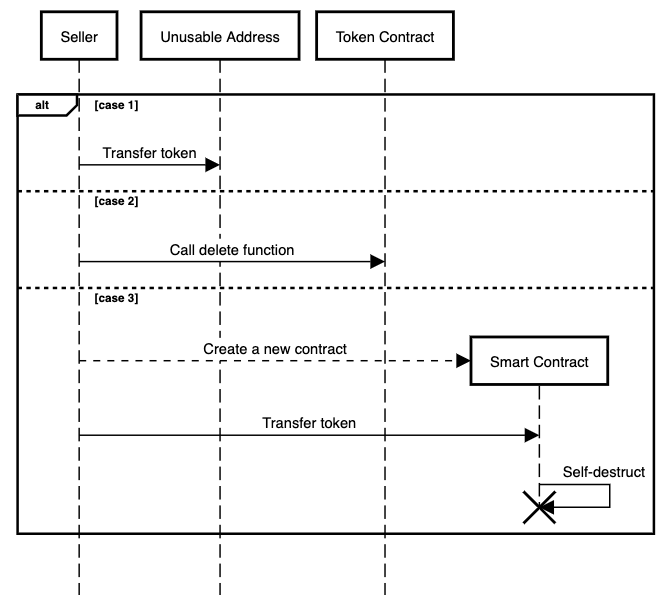}
	\caption{Burned Token.}
	\label{burn_seq}
\end{figure}

\vspace{0.5em}\noindent \textbf{Solution:}
Potential misuse of a token can be prevented by either destroying or locking the token once redeemed (aka., removed from circulation). This process is referred to as \emph{token burning} and could be achieved in several ways. Fig. {\ref{burn_seq}} illustrates three ways of achieving this. 
In the first approach, the seller can transfer the tokens to an unusable/invalid address (aka., burn address). Such an address should not have a corresponding private key that can control the tokens it receives. The second way is to delete a token by calling the delete/burn function on the token contract used to create it or the token registry that keeps track of balances. The third method is to deploy a smart contract that immediately self-destructs as soon as it receives a token. 
For archiving, burned tokens could be moved to the token owner’s read-only token wallet that any party cannot operate on the seller side.

\vspace{0.5em}\noindent \textbf{Consequences:}

\vspace{0.5em}\noindent Benefits:
\begin{itemize}[leftmargin=*]
\item \textit{Redeem once.} Double spending is alleviated as the token is no longer spendable or even exist.
\item \textit{Integrity.} The blockchain ensures the integrity of token status.
\end{itemize}

\vspace{0.5em}\noindent Drawbacks:
\begin{itemize}[leftmargin=*]
 \item \textit{Cost.} If a public blockchain is used, extra cost incurs when calling the smart contract functions to burn the token. 
 \item \textit{Traceability.} The burned tokens cannot be tracked. On the seller side, this can be tracked by moving to the read-only wallet.
\end{itemize}

\vspace{0.5em}\noindent \textbf{Related patterns:} 
Once a token is redeemed, \textit{token template} and \textit{token registry} patterns can use the burned token pattern to mark a token as unspendable.

\vspace{0.5em}\noindent \textbf{Known uses:}
\begin{itemize}[leftmargin=*]
    \item \textit{Perth Mint Gold} token {\cite{Perth_Mint}} holders can redeem their tokens to either get Perth Mint gold or fiat currency. Before redeeming, the token owner needs to call the burn function on the smart contract to burn the token. Once the burned token is confirmed, the mint issues the gold or fiat currency.
    \item \textit{Smart Money}\footnote{See footnote \ref{tietoevry}} developed by TietoEVRY maintains burned tokens in a separate wallet.
    \item In \textit{Making Money Smart}\footnote{See footnote \ref{making_money_smart}} project, sellers can redeem smart tokens for payment in Australian Dollars.
\end{itemize}

\subsection{Seller Credential}
\textbf{Summary:} The buyer can validate the seller's business qualifications through seller credentials.

\vspace{0.5em}\noindent \textbf{Context:}
In blockchain-based payment applications, different policies stipulate rules for the usage of tokens. Some of these rules specify eligible sellers and their products or services regarding their qualifications or qualities. 

\vspace{0.5em}\noindent \textbf{Problem:}
How can the buyer verify the eligibility of the seller before transferring a token to buy a product or use a service? 

\vspace{0.5em}\noindent \textbf{Forces:}
\begin{itemize}[leftmargin=*]
    \item \textit{Immutability.} A payment cannot be reversed if a token is later found to be ineligible to spend for the chosen product, service, or its seller.
    \item \textit{Complexity.} A seller's qualifications or qualities might be complex and difficult to codify.
    \item \textit{Privacy.} The disclosed eligibility information should contain minimum data necessary to identify relevant aspects of oneself without revealing additional information. 
    \item \textit{Authentication.} The authenticity of proof of eligibility documents must be ensured as the documents might be counterfeit. 
\end{itemize}

\begin{figure}[t]
	\centering
	\includegraphics[width=0.85\columnwidth]{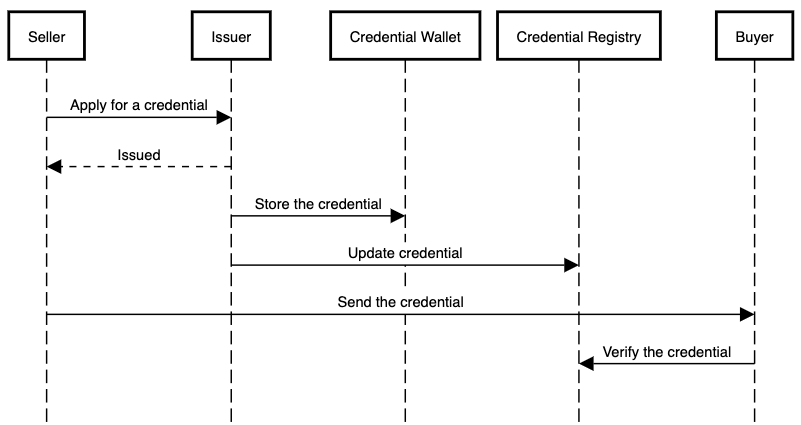}
	\caption{Sequence diagram of seller credential.}
	\label{credential_seq}
\end{figure}

\vspace{0.5em}\noindent \textbf{Solution:} When a token stipulates rules for its usage in term of the eligibility of a seller and its products or services, the sequence of activities illustrated in Fig. {\ref{credential_seq}} can be used to prove the eligibility to a potential buyer. First, the seller applies for a credential from the token issuer. Once approved, the credential is stored in the seller's wallet and the credential registry (i.e., a registry that tracks the seller's attributes using the \textit{token registry} pattern). Third, when a token requires a specific credential from the seller to be validated before spending, the seller can present the credential to the token owner (i.e., buyer) as proof of eligibility. Finally, the buyer then validates the credential from the credential registry. 

A token could stipulate rules for its usage in term of the eligibility of a seller or its products/services using the \textit{policy contract} pattern. Similarly, the seller's credentials, qualifications, and qualities could be specified as a set of attributes (e.g., codified as a JSON schema) or digital certificates signed by the issuer. The credential registry could be implemented similar to the \textit{token registry} pattern. Specific use cases also require credentials of the buyer to be validated by the seller. For example, the seller may want to validate the buyer's age or whether the buyer has the right to spend the given token. In such cases, the pattern can be applied by interchanging the role of seller and buyer. Rather than sending the credential to the seller's wallet, the issuer could send an identifier/link to the credential stored in the credential registry. Either way, the seller should share the credential (or relevant parts of its) or its identifier with the potential buyer. This can be handled off-chain, e.g., using the sharing functionality provided by the seller's wallet or attaching the identifier/link to the invoice. Even if the credential is shared, the buyer may want to check the credential registry to ensure that the issuer has not updated or revoked the credential presented by the seller. 


\vspace{0.5em}\noindent \textbf{Consequences:}

\vspace{0.5em}\noindent Benefits:
\begin{itemize}[leftmargin=*]
\item \textit{Immutability.} As the seller's credential is verified before initiating the payment, need for payment reversal does not arise.
\item \textit{Authenticity.} Digital signatures can guarantee the authenticity of credentials. Also, only the issuer is authorised to add credentials to the credential registry.
\item \textit{Reusability.} The seller credential can be used for all related tokens associated with the same policy.
\item \textit{Cost.} No cost is involved in sharing the credential to multiple buyers as it is handled off-chain.
 \item \textit{Upgradability.} The credential details can be updated once it is issued to a seller by issuing a new credential and updating the credential registry.
\end{itemize}

\vspace{0.5em}\noindent Drawbacks:
\begin{itemize}[leftmargin=*]
 \item \textit{Security.} A seller may present a revoked or old credential; hence, the buyer needs to validate with the credential registry.
 \item \textit{Cost.} A transaction fee needs to be paid on a public blockchain to add a credential to the credential registry. Further, transaction fees need to be paid when the credential verification process involved more complex operations than read-only ledger data access.
\end{itemize}

\vspace{0.5em}\noindent \textbf{Related patterns:} 

The \textit{token template} pattern specifies tokens that require seller validation, and the \textit{policy contract} pattern specifies the seller credentials to be validated. The \textit{token registry} pattern can be used to build the credential registry. When seller credential involves external/physical attributes, those can be verified using the \textit{oracle} pattern.

\vspace{0.5em}\noindent \textbf{Known uses:}
\begin{itemize}[leftmargin=*]
    \item \textit{Smart money}\footnote{See footnote \ref{tietoevry}} developed by TietoEVRY defines spending rules for tokens. Credentials are created and issued to the registered merchants. Credentials need to be checked before making the payment.
    \item \textit{Polymesh}\footnote{See footnote \ref{polymesh}} specify credentials such as seller's KYC status and whether the seller is affiliated, accredited, or exempted from a specific jurisdiction.
    \item \textit{Sovrin}\footnote{\url{https://sovrin.org}} provides a blockchain-based self-sovereign identity infrastructure to issue credentials.
    \item \textit{uPort Serto}\footnote{\url{https://serto.id}} provides a blockchain-based credential management platform to achieve self-sovereign identity.
\end{itemize}

\subsection{Escrow}
\textbf{Summary:} 
Before making a transaction, tokens are transferred to a third-party smart contract called the \textit{escrow}. The escrow holds the deposited tokens until the payment conditions are satisfied. 

\vspace{0.5em}\noindent \textbf{Context:} 
The parties involved in the transaction 
need to ensure that both the agreed product/service is delivered and payment is made. One party should not be able to default the transaction at the expense of the other party.

\vspace{0.5em}\noindent \textbf{Problem:} 
How to ensure that the buyer gets the desired product/service while ensuring the seller gets the payment?

\vspace{0.5em}\noindent \textbf{Forces:} 
\begin{itemize}
    \item \textit{Trust.} It is hard for the parties involved in a transaction to completely trust a third party with their private information and funds.
    \item \textit{Efficiency.} It is time-consuming to establish a trust account and regulate the funds in a centralised payment system.
    \item \textit{Cost.} Centralised payment systems often charge high fees for trust account services. 
    \item \textit{Security.} When the funds are controlled from a single point, there is a possibility of funds could be stolen or mistakes are made.
    \item \textit{Volume.} Buyers and sellers may be engaged in a large number of transactions.
\end{itemize}


\begin{figure}[t]
	\centering
	\includegraphics[width=0.78\columnwidth]{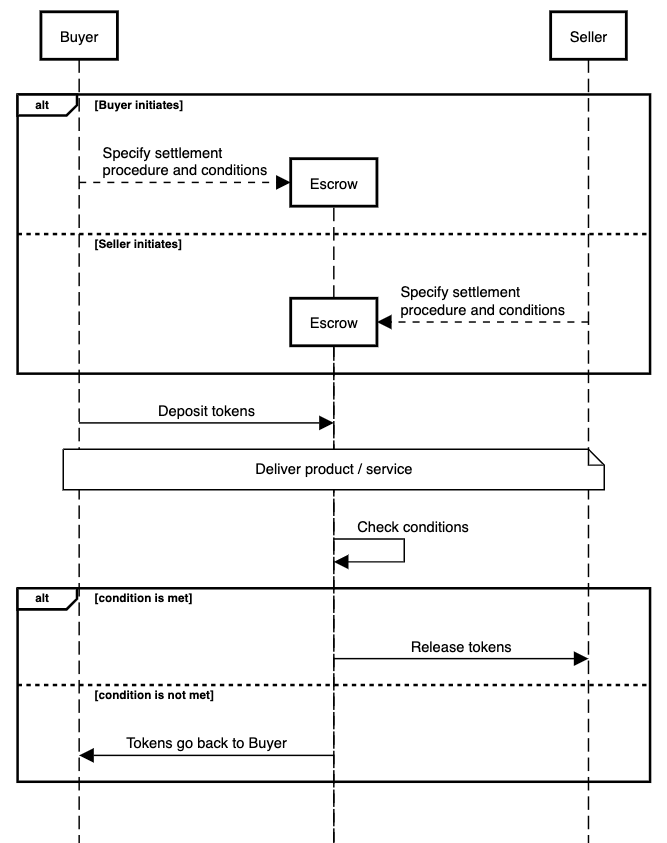}
	\caption{Sequence diagram of escrow.}
	\label{escrow_seq}
\end{figure}

\vspace{0.5em}\noindent \textbf{Solution:} 
As seen in Fig.~\ref{escrow_seq}, a smart contract can play the role of an escrow that holds the fund until the payment conditions are fulfilled. First, specify the settlement procedure and conditions as a smart contract. This smart contract could be specified and deployed by either the seller or buyer. Second, the buyer transfers the token(s) to the escrow smart contract. Third, when token release conditions are met by providing the desired product/server, the respective event is informed to the escrow smart contract. Finally, the escrow validates the pre-defined conditions and releases the tokens to the seller. If the respective event is not informed to the escrow within the stipulated time or the event indicates that the product/service was not delivered as per the agreed terms, then the tokens are sent back to the buyer.

If the payment conditions depend only on on-chain data, then a delegated call can be made to the escrow contract to inform the delivery of the product/service. If the payment conditions depend on external data like the shipment of a product, an \textit{oracle} pattern can be used to provide desired data to the escrow. The security of escrow functionality implemented by the smart contract can be ensured as the smart contract code is immutable once deployed on the blockchain. This gives the parties involved in the transaction confidence that they will not be cheated during the trade. Care must be taken to ensure that the specified settlement conditions are unambiguous. For example, if the delivery is to be performed within 7-days, smart contract should also specify from when clock should start and whether it is sufficient for the seller to ship the product or whether it needs to be received by the buyer by the deadline. Further, it is recommended to specify the time in terms of a block number, as the block timestamp is not precise or could be crafted by a miner within a specific window.

\vspace{0.5em}\noindent \textbf{Consequences:} 
Benefits:
\begin{itemize}
  \item \textit{Trust and security.} An escrow smart contract reduces the risk of fraud by acting as a neutral party and ensuring proper escrow logic execution.
  \item \textit{Transparency.} Operations happening in the system are transparent as relevant transactions are accessible to all blockchain participants.
  \item \textit{Efficiency.} Blockchain eliminates the need for third parties, which in turn helps to reduce transaction cost and enhances service efficiency.
\end{itemize}

Drawbacks: 
\begin{itemize}
    \item \textit{Cost.} Transaction fees need to be paid to deploy and execute escrow smart contract on public blockchains.
    \item \textit{Privacy.} As all blockchain transactions are transparent, escrow transactions can potentially leak sensitive business information, e.g., the rate of disputes with customers. 

\end{itemize}

\vspace{0.5em}\noindent \textbf{Related patterns:} 
Tokens can be transferred to/from the token contract generated using the \textit{token template} pattern. If the \textit{token registry} pattern is used, seller's and buyer's balances need to be updated on the registry. \textit{Oracle} pattern can be used to provide the data required to validate physical product or service delivery. Alternatively, the buyer and seller may sign a transaction using the \textit{multisignature} pattern to inform the escrow about the successful delivery of the product or service. \textit{Stealth address} pattern can be used to generate one-time addresses to transact with the escrow to enhance privacy.

\vspace{0.5em}\noindent \textbf{Known uses:}
\begin{itemize}
  \item \textit{Kleros Escrow}\footnote{\url{https://kleros.io/en/escrow}} is a blockchain-based trustless dispute resolution platform that provides escrow services for cross-chain asset swaps.
  \item \textit{Counos}\footnote{\url{https://counos.io}} is a blockchain platform based in Switzerland, which offers financial and payment services, including \\multisignature-based escrow for cryptocurrencies.
  \item \textit{IBC Group}\footnote{\url{https://ibcgroup.io}} is a blockchain financial services company that offers smart-contract-based cryptocurrency escrow services charging 1\% fee in fiat currency. 
  \item \textit{Merchant Token} {\cite{Merchant_Token}} brings consumer protection concepts from the traditional card payment industry to blockchains by locking away a buyer's payment to a seller with less reputation until delivery of the product/service is confirmed, or disputes are resolved.
\end{itemize}

\subsection{Payment Channel}
\noindent \textbf{Summary:} 
A two-way pathway (aka., payment channel) is established between transacting parties to perform transactions off-chain while recording opening and final settlement transactions on the blockchain.

\vspace{0.5em}\noindent \textbf{Context:} 
A set of low-value payments are to be made frequently, e.g., a small payment paid every time a WiFi service is used. 

\vspace{0.5em}\noindent \textbf{Problem:} 
If a public blockchain is used, micro-payments are too expensive to make as the transaction fee might be higher than the monetary value within the transaction. Also, it takes time to achieve finality on a blockchain. Therefore, how can frequent payments be processed without waiting for a long time and incurring high transaction fees? 

\vspace{0.5em}\noindent \textbf{Forces:} 
\begin{itemize}
  \item \textit{Latency.} It might take a long time to confirm a transaction on the blockchain, although users usually expect instantaneous settlement of payments.
  \item \textit{Cost.} If a public blockchain is used, the transaction fee might be higher than the monetary value of the payment. 
\end{itemize}


\begin{figure}[t]
	\centering
	\includegraphics[width=0.85\columnwidth]{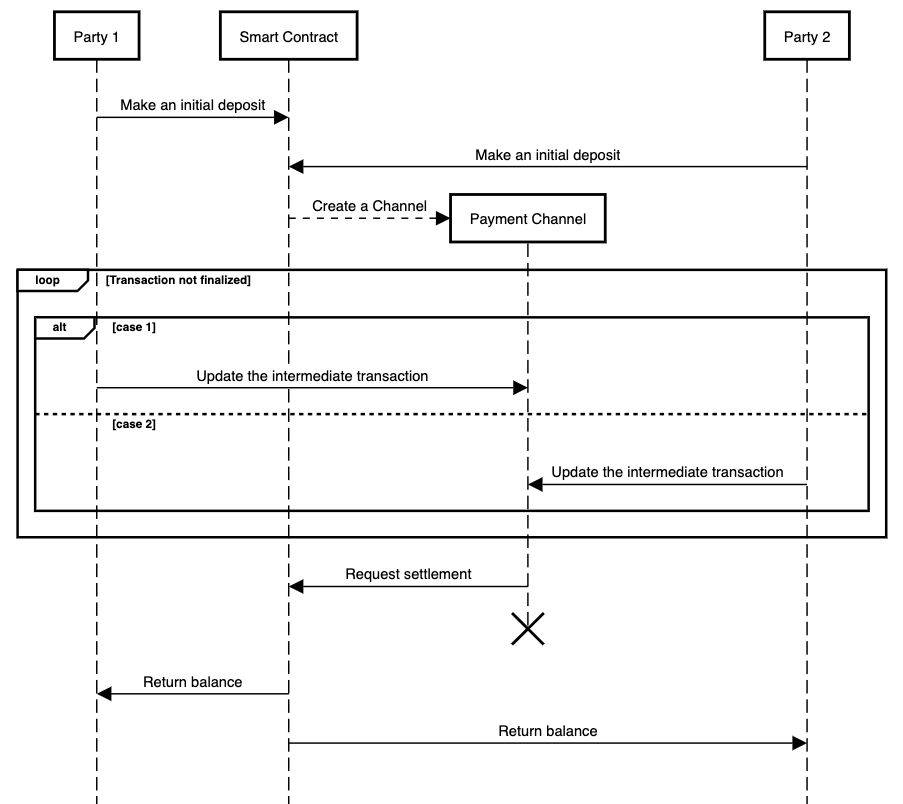}
	\caption{Sequence diagram of payment channel.}
	\label{channel_seq}
\end{figure}

\vspace{0.5em}\noindent \textbf{Solution:} 
An off-chain, two-way pathway (aka., \textit{payment channel}) can be established between two parties to handle frequently occurring transactions. Fig. {\ref{channel_seq}} depicts the sequence of activities involved in a payment channel. First, an initial deposit is made by one or both parties (e.g., buyer and seller) to a smart contract that locks up tokens as a security. The buyer and seller can then engage in a sequence of transactions where both parties digitally sign each intermediate transaction. The signed transactions are recorded off-chain.
When the settlement is to be made, each party's closing balance is informed to the smart contract. Finally, the smart contract returns the tokens as per the closing balance of each party. 

The buyer and seller may also interchange the roles such that tokens flow both ways. Each intermediate transaction should carry a sequence number to correctly calculate the settlement balance and ensure no transaction is missed. Recorded intermediate transactions may also be sent to the smart contract as proof, especially if there is any dispute. Settlement on the blockchain can also be performed periodically depending on the agreement between the two parties. The payment channel is discontinued once the settlement is requested. 

\vspace{0.5em}\noindent \textbf{Consequences:} 

Benefits:
\begin{itemize}
  \item \textit{Performance.} The off-chain micro-payment transactions can be settled instantaneously without waiting for finality on the blockchain.
  \item \textit{Throughput.} Off-chain payment channels can achieve much higher throughput as they are not constrained by inter-block time.
  \item \textit{Privacy.} Intermediate transactions processed by payment channels are not publicly available. 
  \item \textit{Cost.} If a public blockchain is used, only the initial deposit and final settlement transactions incur transaction fees.
\end{itemize}

Drawbacks: 
\begin{itemize}
   \item \textit{Data Integrity.} The data integrity of the intermediate state of payment channels cannot be ensured because intermediate transactions are not stored on-chain and vulnerable to transaction malleability. However, disputes can be usually resolved by presenting all the intermediate transactions to the smart contract as evidence.
   \item \textit{Lack of Liquidity.} To establish a payment channel, tokens from one or both parties of the channel needs to be locked in the smart contract during the payment channel's lifecycle. 
\end{itemize}

\vspace{0.5em}\noindent \textbf{Related patterns:} \textit{Token registry} pattern could be used by the smart contract to update the balance of the transacting parties at the time of payment channel initiation and settlement.

\vspace{0.5em}\noindent \textbf{Known uses:}
\begin{itemize}
  \item \textit{Bitcoin Lightning Network}\footnote{\url{https://blockstream.com/lightning/}} adopts an off-chain protocol to enable micro-payments. It only broadcasts the final version of the funding transaction to settle the payment in the Bitcoin network. 

  \item \textit{Ethereum Raiden Network}\footnote{\url{https://raiden.network/}} uses a network of off-chain payment channels that enables secure money transfer.
  
  \item \textit{Ethereum State Channel}\footnote{\url{http://jeffcoleman.ca/state-channels/}} provides a channel protocol that supports exchanging state for general-purpose blockchain applications. 

\end{itemize}

\subsection{Stealth Address}
\textbf{Summary: } 
Use a one-time address (aka., stealth address) for the transaction to protect the the privacy of the parties involved in the payment. 

\vspace{0.5em}\noindent \textbf{Context:}
Sellers and buyers are engaged in multiple transactions and are concerned about their privacy. 

\vspace{0.5em}\noindent \textbf{Problem:}
How can transacting parties protect their privacy when engaging in a large number of transactions?
 
\vspace{0.5em}\noindent \textbf{Forces:}
\begin{itemize}[leftmargin=*]
    \item \textit{Privacy.} While a blockchain transaction is pseudonymous, it is possible to correlate transactions and reveal the real-world identity by linking transactions with other data (e.g., IP address). For example, by analysing the transaction graph, flow of tokens, and other external information, third parties could correlate the transactions and observe buyers' and sellers' behaviour and identify the entity behind an address.
    \item \textit{Volume.} Buyers and sellers may engage in a large number of transactions.
    \item \textit{Traceability.} Transacting parties and regulators expect to track the flow of tokens.
    \item \textit{Auxiliary information.} Depending on the use case, buyers' and sellers' information may be available on-chain (e.g., \textit{seller credential}) or off-chain (e.g., trade reports and census data).
\end{itemize}


\begin{figure}[t]
	\centering
	\includegraphics[width=0.7\columnwidth]{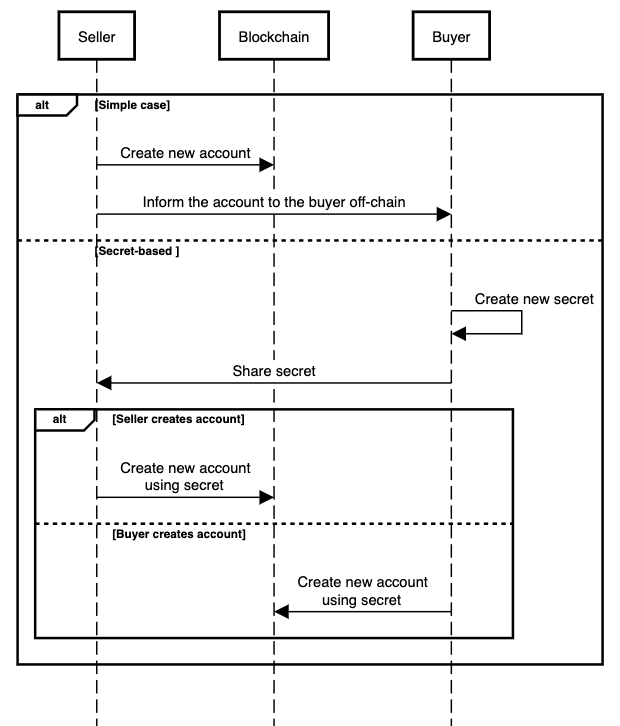}
	\caption{Sequence diagram of stealth address.}
	\label{stealth_seq}
\end{figure}

\vspace{0.5em}\noindent \textbf{Solution:} Buyer's and seller's anonymity can be enhanced by using a one-time address referred to as the \textit{stealth address}. Fig. {\ref{stealth_seq}} illustrates the sequence of activities required to generate a stealth address and share it between transacting parties using two techniques. Note that the buyer and seller roles in Fig. {\ref{stealth_seq}} are interchangeable. 
In the first technique, the seller creates a one-time address/account for each transaction on behalf of the buyer such that different transactions sent to the same recipient are unlinkable. The seller could inform the new account to the buyer off-chain, e.g., an invoice can contain the new address. The second technique shares a key/secret off-chain between the two parties that can be used to generate a new account for each future transaction. For example, a one-time secret can be shared between the parties to use as the seed to create a public-private key pair of a new account. Then the hash of that key can be used as a seed to create the next account. This could be extended to a hash chain, where each round of hashing produces a datum that can be used as the seed to create a new public-private key pair. However, this would mean both seller and buyer can spend the token as the private key for the account are known to both. This could be overcome by modifying the public-private key generation process as presented in the BIP-32 standard\footnote{\url{https://github.com/bitcoin/bips/blob/master/bip-0032.mediawiki}}. A couple of other solutions could also be used to share the secret on-chain, e.g., reuse a public key, reuse a signature nonce, and the seller's addressing requiring to have a specific prefix\footnote{\url{https://sourceforge.net/p/bitcoin/mailman/message/31813471/}}.


\vspace{0.5em}\noindent \textbf{Consequences:}

\vspace{0.5em}\noindent Benefits:
\begin{itemize}[leftmargin=*]
\item \textit{Privacy.} As a new address is used for each transaction, it is not straightforward to correlate and trace transactions on the blockchain. This enhances the protection of the real-world identity of the recipient.
\end{itemize}

\vspace{0.5em}\noindent Drawbacks:
\begin{itemize}[leftmargin=*]
 \item \textit{Traceability.} As new addresses are created based on an off-chain secret, it is difficult to track transactions targeted to the same physical recipient.
 Such a lack of traceability could be leveraged to engage in illegal activities. 
 \item \textit{Volume.} The recipient of payments will find it difficult to link, reconcile, redeem, and manage transactions as each transaction is linked to a different address. Moreover, transactions may still contain additional details such a product/service types and quantities, when aggregated, could reveal transacting parties behaviour and identity. 
 Furthermore, if the recipient wants to later aggregate tokens without revealing its identity, it has to use token mixing services, which incur additional costs. 
\end{itemize}

\vspace{0.5em}\noindent \textbf{Related patterns:} Stealth addresses can be used with \textit{token registry} and \textit{escrow} patterns to hide the address of sellers and buyers.

\vspace{0.5em}\noindent \textbf{Known uses:}
\begin{itemize}[leftmargin=*]
\item \textit{Basic Stealth Address Protocol} (BSAP)\footnote{\url{https://lists.linuxfoundation.org/pipermail/bitcoin-dev/2014-January/004020.html}} adopts the Elliptic Curve Diffie-Hellman (ECDH) protocol to deal with Bitcoin transactions. \textit{Improved Stealth Address Protocol} (ISAP)\footnote{\url{https://bytecoin.org/old/whitepaper.pdf}} is an upgraded version of BSAP with an additional key creation feature. The recipient is the only entity that can compute the private key for the temporary address to receive Bitcoin.
\item To address overuse of private spending key, Dual-Key Stealth Address Protocol (DKSAP) is designed for a wallet solution \textit{ShadowSend}\footnote{\url{https://github.com/shadowproject}} uses two pairs of keys, a scan key pair and a spend key pair to provide decentralised anonymous currency.

\end{itemize}

\subsection{Oracle}
\textbf{Summary: } To examine the fulfilment of payment conditions, an oracle is used to introduce the state of external systems into the blockchain execution environment.

\vspace{0.5em}\noindent \textbf{Context:}
The payment is conditional on the physical properties of a product/service or completion of a real-world action, e.g., quality attributes of a product or shipment of a product. Typically such data are provided or certified by third parties.

\vspace{0.5em}\noindent \textbf{Problem:} 
How to fetch external state/data into a smart contract to decide on a payment condition?

\vspace{0.5em}\noindent \textbf{Forces:} 
\begin{itemize}
  \item \textit{Connectivity.} Blockchain-based payment applications may require external state data, e.g., the ambient temperature of a container. However, blockchain is a self-contained execution environment. Smart contracts in blockchain-based payment applications cannot read the state data of external devices or systems.
  \item \textit{Trust.} Both the transacting parties need to agree on the state of a product and its attributes. When such data are provided by a third party, the provider needs to be trustable to transacting parties, including any regulators.
  \item \textit{Availability.} The external devices or systems that provide data may change or even disappear.
\end{itemize}


\begin{figure}[t]
	\centering
	\includegraphics[width=0.7\columnwidth]{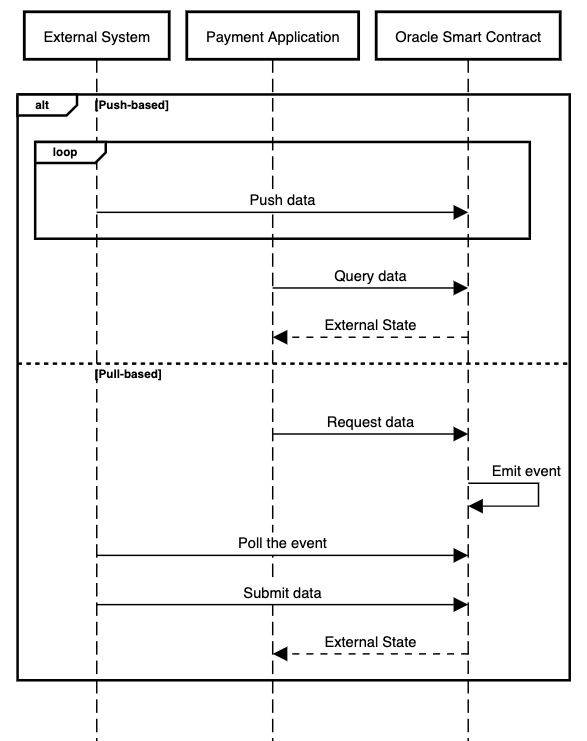}
	\caption{Sequence diagram of oracle.}
	\label{oracle_seq}
\end{figure}

\vspace{0.5em}\noindent \textbf{Solution:} 
A third-party hosted smart contract can be used to provide data to the blockchain from external servers, websites, and IoT devices. These external devices/systems call functions on the smart contract to record the data on the blockchain using their private key attesting the data origin. The system consisting of the smart contract and external elements are referred to as an \textit{oracle}. As illustrated in Fig. {\ref{oracle_seq}}, an oracle could allow other smart contracts to either push or pull data from off-chain components. In the first case, the external system sends the desired data to the oracle smart contract using a transaction. This transaction should invoke a method that accepts data as an input parameter. When the payment application needs to access the data, it can make a delegated call to the oracle smart contract to read the data. In the second case, the payment application indicates its need to access external data to the oracle using a delegated call. In response, the oracle emits a blockchain event indicating the desired data. External system should poll for such events to determine what data are requested by the oracle. As the blockchain cannot directly interact with external systems, events provide a passive communication mechanism between a smart contract and external systems that poll for relevant events. Once the requested data are available, the external application submits the data to the blockchain using a transaction. Consequently, the oracle makes a delegated call to the smart contract to deliver the requested data.

When the smart contract calls the oracle, if it does not have the desired or latest data, a subscription could be made to the oracle by specifying the smart contract address and method to be called. Once the data are available, the oracle can call the smart contract using a delegated call. A new transaction could be issued every time the external state changes. Even if there is no change in data, some applications may require the external system to submit a transaction indicating no change in data. 
There are two types of oracles: centralised and decentralised oracles. A centralised oracle introduces a single trusted third party into the system, which might become a single point of failure. Alternatively, a decentralised oracle is designed based on multiple external servers and multiple independent data sources, overcoming the single point of failure. A decentralised oracle can use multisignature or voting to reach a consensus on the validity of data injected into the oracle. Similarly, at the hardware level, redundant IoT devices could be deployed to deal with adversary IoT devices that produce malicious data. Then the oracle smart contract can cross-check the consistency of data collected by different IoT devices.

Oracles can be used to access the physical attributes of a product to avoid frauds in trading. A physical item's integrity can be verified by connecting the physical item with its digital representation (i.e., \textit{digital-physical parity}). The physical attributes of an item (e.g., chemical composition, geo-location, and visual features) can be extracted via various techniques such as genomics analysis, isotope analysis, and machine learning. The captured attributes can be digitally recorded on the blockchain through oracles to ensure immutability and transparency.

\vspace{0.5em}\noindent \textbf{Consequences:} 

Benefits:
\begin{itemize}
  \item \textit{Connectivity.} Payment application's smart contracts can access external states through the oracle.
  \item \textit{Trust.} A decentralised oracle increases trust as it relies on trust on a majority of third parties. 
  \item \textit{Availability} A decentralised oracle can remove the risk of a single point of failure and increase availability.

\end{itemize}

Drawbacks: 
\begin{itemize}
   \item \textit{Trust.} A centralised oracle is a third party integrated into the system, which needs to be trusted by all stakeholders involved in the payment. 
   \item \textit{Data Integrity.} Validators can only check the digital signature of the oracle but cannot check the correctness or originality of the injected external data.
   \item \textit{
   Availability} A centralised oracle may lead to a single point of failure. 
   \item \textit{Cost.} There is a cost in injecting data into oracle smart contract and looking up data from the oracle. Further, the cost of injecting data from the external world increases with the number of oracles being used and the rate of data change.
  \item \textit{Time.} It could take a longer time for the decentralised oracle to reach a consensus over the state data to be injected. 
\end{itemize}

\vspace{0.5em}\noindent \textbf{Related patterns:} \textit{Policy contract} and \textit{seller credential} patterns can use an oracle to verify physical attributes about a seller or its products. The \textit{multisignature} pattern can be used to sign the data submitted to a decentralised oracle.

\vspace{0.5em}\noindent \textbf{Known uses:}
\begin{itemize}
  \item \emph{Oracle} in Bitcoin\footnote{\url{https://en.bitcoin.it/wiki/Contract\#Example_4:_Using_external_state}} is a server outside the Bitcoin blockchain, which can evaluate user-defined condition expressions for payment based on the external state data.
  \item \emph{Provable}\footnote{\url{https://provable.xyz/}} is an oracle provider, which utilises trusted hardware to directly collect data from the external environment, which can be used to check payment conditions. 
  \item \textit{Orisi}\footnote{\url{http://orisi.org/}} is an oracle framework for Bitcoin smart contracts that enables the transaction participants to select a set of oracles and apply multisignature to confirm the status.
  \item \textit{Merchant Token} {\cite{Merchant_Token}} uses a decentralised oracle to confirm a product/service delivery or resolve disputes before releasing funds locked in the escrow to the seller.
  
\end{itemize}

\subsection{Multisignature}

\noindent \textbf{Summary:} 
Given a pool of $n$ addresses that could authorise a payment, get a subset $m$ of them to authorise a payment by signing the respective transaction ($m \leq n$). 

\vspace{0.5em}\noindent \textbf{Context:} 
A transaction needs to be authorised by multiple parties (i.e., blockchain addresses), e.g., to release tokens held by an escrow, both buyer and seller need to confirm successful delivery of product or service.

\vspace{0.5em}\noindent \textbf{Problem:} How can more than one party approve a transaction? 

\vspace{0.5em}\noindent \textbf{Forces:} 
\begin{itemize}
  \item \textit{Dynamism.} A transaction needs to be authorised by a subset of authorities pre-defined in a pool of addresses. However, some of the authorities may not be available, unreachable, or related private key may be compromised when issuing the transaction. Also, the address pool needs to be changed as the payment application evolves, e.g., the addition of a new authority or revocation of an existing authority.
  \item \textit{Tolerance of lost key.} Losing a private key on the blockchain means permanent loss of control over a blockchain account and associated smart contracts as blockchain does not provide any key recovery mechanism. 
\end{itemize}

\begin{figure}[t]
	\centering
	\includegraphics[width=0.3\columnwidth]{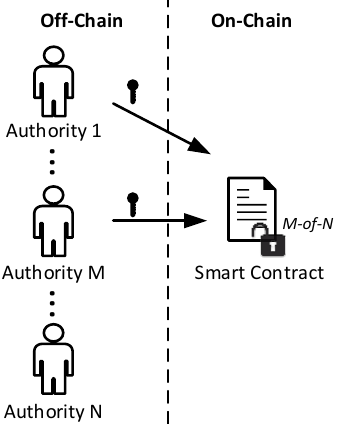}
	\caption{Multisignature.}
	\label{multisignature}
\end{figure}


\vspace{0.5em}\noindent \textbf{Solution:} 
A transaction could indicate multi-party authorisation using a digital signature scheme that allows a group of addresses to sign the transaction. Such a signature is known as a \textit{multisignature}. Depending on the chosen multisignature algorithm, we can enforce that the transaction is signed by all or a subset of the authorised addresses. As depicted in Fig.~\ref{multisignature}, to enable flexible binding of pre-defined authorities for a payment transaction, a multisignature mechanism can be designed to require $m$ out of $n$ private keys to authorise a transaction, in which $m$ is the threshold of authorisation. 

Depending on the context, either the token issuer or buyer can pre-define a group of addresses that can authorise a payment and set the minimal number of authorisations required to approve a payment. While some blockchain platforms accept transactions containing a multisignature, a multisignature is more commonly validated using smart contracts. In the simplest form, each authorised address can submit a signed transaction to the multisignature smart contract that keeps track of the number of transactions with valid signatures. Once the number of valid transactions reaches the pre-specified threshold, the smart contract can perform the token payment. Alternatively, more efficient signature schemes, such as threshold signatures, could be used to generate a short signature off-chain and submit to the smart contract using a single transaction. In either case, the pool of authorised addresses needs to be pre-specified before issuing the transaction. When the membership change, threshold signature schemes usually require regeneration and distribution of all the keys/addresses. 

\vspace{0.5em}\noindent \textbf{Consequences:} 

Benefits:
\begin{itemize}
  \item \textit{Dynamism.} Multisignature allows flexible binding of authorities for approving a transaction. 
  \item \textit{Tolerance of lost keys.} The risk of losing control over smart contracts is reduced as one participant can have more than one blockchain address in case a private key is lost. 
\end{itemize}

Drawbacks: 
\begin{itemize}
  \item \textit{Risk of losing keys.} There is still a risk of losing control over smart contracts when $m$ private keys among the $n$ private keys are lost.
  \item \textit{Cost.} If a public blockchain is adopted, storing multiple addresses, updating the list of addresses, and issuing multiple transactions cost money. 
\end{itemize}

\vspace{0.5em}\noindent \textbf{Related patterns:} The \textit{escrow} pattern can use a multisignature to get confirmation from both buyer and seller about the successful delivery of a product. A decentralised \textit{oracle} can use multisignature to confirm the physical state of a product from multiple attesters. 

\vspace{0.5em}\noindent \textbf{Known uses:}
\begin{itemize}
  \item \textit{MultiSignature}\footnote{\url{https://en.bitcoin.it/wiki/Multisignature}} is offered by Bitcoin to authorise transactions.
  \item Multisignature wallet is provided in the \textit{Ethereum Mist DApp} browser\footnote{\url{https://github.com/ethereum/mist}} to approve transactions.
  \item A multisignature wallet is used in \textit{TokenMarket} security token to confirm critical activities that affect the shareholder balances.
  \item Data filed in the \textit{ERC-1400}\footnote{See footnote \ref{erc_1400}} standard can be used to submit a multisignature signed off-chain using a single transaction. 
\end{itemize}

\subsection{Token Swap}
\noindent \textbf{Summary:}
Token swap allows users to trade directly between two types of tokens as an atomic transaction.
 
\vspace{0.5em}\noindent \textbf{Context:} 
A blockchain-based payment application supports different types of tokens, e.g., different asset classes.

\vspace{0.5em}\noindent \textbf{Problem:} 
How can users buy and sell tokens for other types of tokens without the risk of other-party not transferring their tokens?

\vspace{0.5em}\noindent \textbf{Forces:} 
\begin{itemize}
  \item \textit{Liquidity.} Users need to convert tokens to other types of tokens.
  \item \textit{Atomicity.} Exchange of token should be atomic based on a rate/amount agreed by both parties of the transaction. 
  \item \textit{Trust.} It is hard for the parties involved in a transaction to trust each other completely.
\end{itemize}


\begin{figure}[t]
	\centering
	\includegraphics[width=0.69\columnwidth]{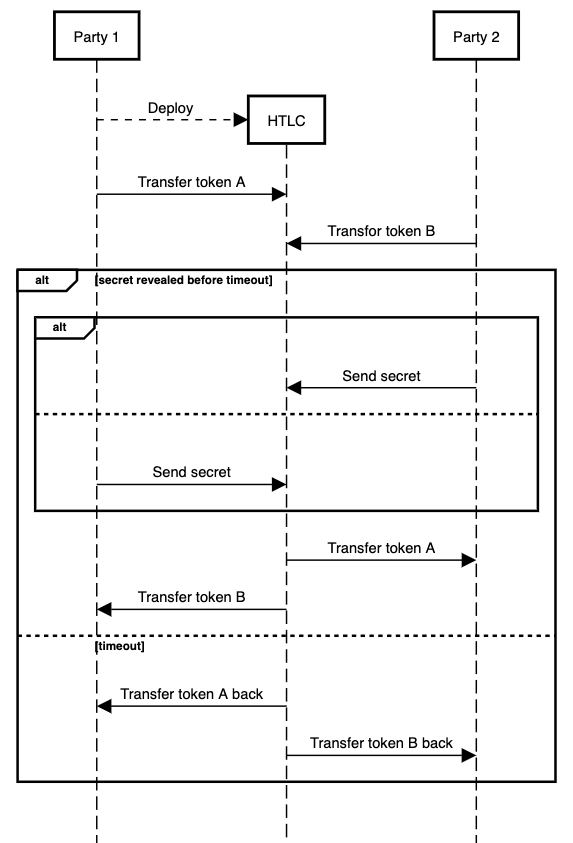}
	\caption{Sequence diagram of token swap.}
	\label{swap_seq}
\end{figure}

\vspace{0.5em}\noindent \textbf{Solution:} By making the token swap atomic, we can ensure both parties will either receive the respective tokens or none of the tokens. A \textit{token swap} is an agreement between two parties that exchange different token types (say token $A$ and token $B$). In a token swap, one party will pay a certain amount of token $A$ to the other party and receive the agreed amount of token $B$ in return. The token swap process is usually managed using a Hashed Timelock Contract (HTLC) deployed on the blockchain. Multiple HTLCs are used to swap tokens in payment channels and across blockchains with the help of an application-layer integrator. Fig.~\ref{swap_seq} illustrates the steps involved in atomically swapping two tokens using a HTLC. First, a HTLC smart contract is deployed by one of the parties on-chain. HTLC specifies a hash generated from an off-chain secret and a timeout specified as a block number. Second, each party transfers the agreed amount of tokens to the HTLC. For example, party 1 transfers token $A$ while party 2 transfers token $B$. Next, to swap the tokens, the off-chain secret is sent to the HTLC using a transaction. Finally, if the revealed off-chain secret generates the same hash embedded in the HTLC, the contract exchanges tokens by transfers token $A$ to party 2 and token $B$ to party 1. If the off-chain secret is not revealed before the timeout, tokens are sent back to their owners.

In payment channels and inter-blockchain token swap, two or more intermediate HTLC contacts are deployed, forming a chain of HTLCs. While each HTLC specifies the same hash generated from the off-chain secret, timeout decreases as we move along the HTLC chain. Then the off-chain secret is revealed from the end of the HTLC chain, leading to a chain reaction that enables each HTLC to swap the tokens it holds. In practice, one HTLC does not automatically transfer the off-chain secret to the next HTLC on the chain. Instead, each party interested in receiving tokens locked by a particular HTLC continues to look for a transaction that reveals the off-chain secret. Once it is observed, the off-chain secret is extracted from the transaction and submitted in a new transaction to the respective HTLC to unlock the tokens. The off-chain secret must be revealed to a HTLC before its timeout. 

\vspace{0.5em}\noindent \textbf{Consequences:} 

Benefits:
\begin{itemize}
  \item \textit{Liquidity.} A user can use his/her tokens to buy other types of tokens, increasing liquidity.
  \item \textit{Data Integrity.} The data integrity of the swapped tokens can be ensured because the token swap process and respective transactions are stored on-chain. 
  \item \textit{Atomicity.} Smart contracts guarantee the atomicity of token swap.
  \item \textit{Cost.} As smart contracts can manage the token swap process, no third party service fee is incurred, e.g., payment to an escrow. 
  \item \textit{Interoperability.} Interoperability is increased through cross-chain token swap.
\end{itemize}

Drawbacks: 
\begin{itemize}
  \item \textit{Privacy.} Token swap transactions are publicly visible. 
  \item \textit{Cost.} There might be additional cost due to the exchanged rate. Also, if a public blockchain is used, there is a cost to deploy and use HTLC smart contract.
  \item \text{Lack of flexibility.} If a party does not withdraw tokens out in time by submitting the off-chain secret, the deposited tokens will be locked in the HTLC or sent back to the payer.
\end{itemize}

\vspace{0.5em}\noindent \textbf{Related patterns:} When exchanging on-chain assets like tokens, this pattern could be used over the \textit{escrow} pattern, as no external confirmation of delivery of assets is needed. The \textit{token registry} pattern can be used to send/receive tokens to/from HTLC.

\vspace{0.5em}\noindent \textbf{Known uses:}
\begin{itemize}
  \item \textit{Metamask}\footnote{\url{https://metamask.io/}} supports token swap feature to compare and swap tokens directly within the wallet and browsers. 
  \item \textit{Kaileido}\footnote{\url{https://www.kaleido.io/blockchain-platform/token-swap}} offers a simple process to swap tokens using HTLCs securely.
  \item The \textit{Bitcoin Lightning Network} {\cite{Lightning_Network}} supports multi-hop payment channels by linking existing payment channels through a chain of HTLCs. This enables parties to transact beyond their payment channel.
  \item \textit{AirSwap}\footnote{\url{https://www.airswap.io/}} supports atomic token swap with a guaranteed price. It also supports swap via Metamask wallet and custom swaps with parties trading tokens.
\end{itemize}


\subsection{Authorised Spender}
\noindent \textbf{Summary:} An authorised third-party spender can spend a certain amount of tokens owned by the approver/buyer.

\vspace{0.5em}\noindent \textbf{Context:} In blockchain applications, buyers need to transfer tokens to sellers to buy products or services.

\vspace{0.5em}\noindent \textbf{Problem:} How can a buyer delegate a third party to pay on behalf of him/her?

\vspace{0.5em}\noindent \textbf{Forces:} 
\begin{itemize}
  \item \textit{Security.} The approval process needs to be secure in case there is a dishonest delegate. For example, the delegate should be able to spend only within the approved allowance.
  \item \textit{Complexity.} It is usually difficult to delegate a third party to pay on behalf of token owners. 
\end{itemize}

\begin{figure}[t]
	\centering
	\includegraphics[width=0.7\columnwidth]{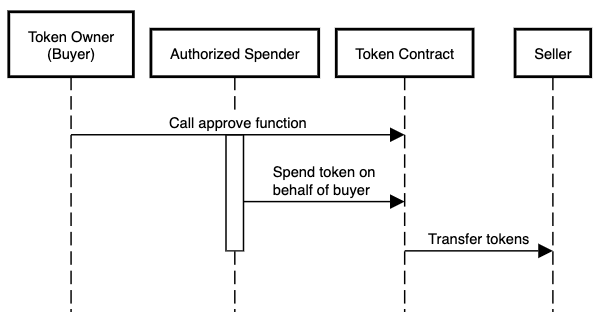}
	\caption{Sequence diagram of authorised spender.}
	\label{spender_seq}
\end{figure}

\vspace{0.5em}\noindent \textbf{Solution:} 
As seen in Fig. {\ref{spender_seq}}, a token owner can approve a third-party user or a smart contract to transfer a certain number of tokens on his/her behalf. First, an approval step can be designed to allow the spenders to use a certain amount of tokens. For example, an approve method can be added to the token template contract with third-party spender address and allowance as parameters. Alternatively, an authorised spender and allowance can be specified on the token registry. Next, when the third party needs to perform a payment on behalf of the buyer, it can send a transaction to the token contract requesting tokens to be transferred from the buyer to the sellers updating the balance of both accounts. Simultaneously, approved balance needs to be updated.

\vspace{0.5em}\noindent \textbf{Consequences:} 

Benefits:

\begin{itemize}
  \item \textit{Flexibility.} Buyers can find a delegate to pay on their behalf within the allocated allowance. 
  \item \textit{Automation.} The buyers can automate payments by relying on the approved third party to perform the payment.
  \item \textit{Security.} The tokens are transferred securely according to the conditions predefined in smart contracts.
\end{itemize}

Drawbacks: 
\begin{itemize}
  \item \textit{Cost.} If a public blockchain is used, an additional cost is incurred to call the approval function.
  \item \textit{Security.} There might be a bug in the smart contract that results in unexpected transfers by the deligate (e.g., wrong address or spending beyond the allowance). 
\end{itemize}

\vspace{0.5em}\noindent \textbf{Related patterns:} The \textit{token template} and \textit{token registry} patterns are needed to support setting an authorised spender and an allowance for the delegate to act on behalf of the buyer.

\vspace{0.5em}\noindent \textbf{Known uses:}
\begin{itemize}
  \item Both \textit{ERC-20}\footnote{See footnote \ref{erc_20}} and \textit{ERC-721}\footnote{See footnote\ref{erc_721}} use an approve function to enable a third party or a smart contract to transfer up to a specified number of tokens. 
  \item \textit{Ethex}\footnote{\url{https://ethex.market}} supports setting a custom allowance for token trading to ensure the decentralised exchange cannot spend more than the allowance even if its smart contract is compromised.
  \item \textit{Saturn Network}\footnote{\url{https://www.saturn.network}} offers an approve feature that allows users to submit a transaction to approve Saturn Network to trade tokens on their behalf.
\end{itemize}

\section{Conclusion}
\label{conclusion}

To systematically organise knowledge on designing payment applications using blockchain, we collected 12 patterns covering critical aspects of blockchain-based payment applications. We identify the lifecycle of a token, in which the state transitions are associated with the collected patterns. The lifecycle along with the annotated patterns, provide a payment-focused systematic view of system interactions and a guide to the effective usage of the patterns in software architecture design. In future, we plan to extend our decision model to select blockchain-based patterns \cite{decision_model} to include payment patterns.

\input{main.bbl}


\end{document}

%% file: main.bbl

%% file: main.bbl
\begin{thebibliography}{10}
\providecommand{\url}[1]{#1}
\csname url@samestyle\endcsname
\providecommand{\newblock}{\relax}
\providecommand{\bibinfo}[2]{#2}
\providecommand{\BIBentrySTDinterwordspacing}{\spaceskip=0pt\relax}
\providecommand{\BIBentryALTinterwordstretchfactor}{4}
\providecommand{\BIBentryALTinterwordspacing}{\spaceskip=\fontdimen2\font plus
\BIBentryALTinterwordstretchfactor\fontdimen3\font minus
  \fontdimen4\font\relax}
\providecommand{\BIBforeignlanguage}[2]{{%
\expandafter\ifx\csname l@#1\endcsname\relax
\typeout{** WARNING: IEEEtran.bst: No hyphenation pattern has been}%
\typeout{** loaded for the language `#1'. Using the pattern for}%
\typeout{** the default language instead.}%
\else
\language=\csname l@#1\endcsname
\fi
#2}}
\providecommand{\BIBdecl}{\relax}
\BIBdecl

\bibitem{2019-Bratanova-ACS}
\BIBentryALTinterwordspacing
A.~Bratanova \emph{et~al.}, ``Blockchain 2030: {A} look at the future of
  blockchain in {Australia},'' Data61, CSIRO, Brisbane, Australia, Tech. Rep.,
  Apr. 2019. [Online]. Available:
  \url{https://www.researchgate.net/publication/332298704_Blockchain_2030_A_Look_at_the_Future_of_Blockchain_in_Australia}
\BIBentrySTDinterwordspacing

\bibitem{xu2018pattern}
X.~Xu, C.~Pautasso, L.~Zhu, Q.~Lu, and I.~Weber, ``A pattern collection for
  blockchain-based applications,'' in \emph{23rd European Conf. on Pattern
  Languages of Programs}, 2018, pp. 1--20.

\bibitem{eberhardt2017or}
J.~Eberhardt and S.~Tai, ``On or off the blockchain? {I}nsights on off-chaining
  computation and data,'' in \emph{European Conf. on Service-Oriented and Cloud
  Computing}.\hskip 1em plus 0.5em minus 0.4em\relax Springer, 2017, pp. 3--15.

\bibitem{smartContractICBC}
Y.~Liu, Q.~Lu, X.~Xu, L.~Zhu, and H.~Yao, ``Applying design patterns in smart
  contracts,'' in \emph{Blockchain -- ICBC 2018}, S.~Chen, H.~Wang, and L.-J.
  Zhang, Eds.\hskip 1em plus 0.5em minus 0.4em\relax Cham: Springer
  International Publishing, 2018, pp. 92--106.

\bibitem{wohrer2018smart}
M.~Wohrer and U.~Zdun, ``Smart contracts: {S}ecurity patterns in the {E}thereum
  ecosystem and {S}olidity,'' in \emph{2018 Int. Workshop on Blockchain
  Oriented Software Engineering (IWBOSE)}.\hskip 1em plus 0.5em minus
  0.4em\relax IEEE, 2018, pp. 2--8.

\bibitem{SSIpattern}
Y.~Liu, Q.~Lu, H.-Y. Paik, and X.~Xu, ``Design patterns for blockchain-based
  self-sovereign identity,'' in \emph{25th European Conf. on Pattern Languages
  of Programs (EuroPLoP'20)}, 2020.

\bibitem{data_migration}
\BIBentryALTinterwordspacing
H.~D. Bandara, X.~Xu, and I.~Weber, ``Patterns for blockchain data migration,''
  in \emph{Proceedings of the European Conference on Pattern Languages of
  Programs 2020}, ser. EuroPLoP '20.\hskip 1em plus 0.5em minus 0.4em\relax New
  York, NY, USA: Association for Computing Machinery, 2020. [Online].
  Available: \url{https://doi.org/10.1145/3424771.3424796}
\BIBentrySTDinterwordspacing

\bibitem{zhang2017applying}
P.~Zhang, J.~White, D.~C. Schmidt, and G.~Lenz, ``Applying software patterns to
  address interoperability in blockchain-based healthcare apps,'' \emph{arXiv
  preprint arXiv:1706.03700}, 2017.

\bibitem{oracle}
R.~M{\"u}hlberger, S.~Bachhofner, E.~Castell{\'o}~Ferrer, C.~Di~Ciccio,
  I.~Weber, M.~W{\"o}hrer, and U.~Zdun, ``Foundational oracle patterns:
  Connecting blockchain to the off-chain world,'' in \emph{Business Process
  Management: Blockchain and Robotic Process Automation Forum}, A.~Asatiani,
  J.~M. Garc{\'i}a, N.~Helander, A.~Jim{\'e}nez-Ram{\'i}rez, A.~Koschmider,
  J.~Mendling, G.~Meroni, and H.~A. Reijers, Eds.\hskip 1em plus 0.5em minus
  0.4em\relax Cham: Springer International Publishing, 2020, pp. 35--51.

\bibitem{bartoletti2017empirical}
M.~Bartoletti and L.~Pompianu, ``An empirical analysis of smart contracts:
  {P}latforms, applications, and design patterns,'' in \emph{Int. Conf. on
  Financial Cryptography and Data Security}.\hskip 1em plus 0.5em minus
  0.4em\relax Springer, 2017, pp. 494--509.

\bibitem{Wohrer_Zdun}
M.~{Wöhrer} and U.~{Zdun}, ``Design patterns for smart contracts in the
  ethereum ecosystem,'' in \emph{2018 IEEE International Conference on Internet
  of Things (iThings) and IEEE Green Computing and Communications (GreenCom)
  and IEEE Cyber, Physical and Social Computing (CPSCom) and IEEE Smart Data
  (SmartData)}, 2018, pp. 1513--1520.

\bibitem{IEEESoftware2017}
Q.~Lu and X.~Xu, ``Adaptable blockchain-based systems: A case study for product
  traceability,'' \emph{IEEE Software}, vol.~34, no.~6, pp. 21--27, November
  2017.

\bibitem{originChain}
X.~Xu, Q.~Lu, Y.~Liu, L.~Zhu, H.~Yao, and A.~V. Vasilakos, ``Designing
  blockchain-based applications a case study for imported product
  traceability,'' \emph{Future Generation Computer Systems}, vol.~92, pp.
  399--406, 2019.

\bibitem{ubaas}
Q.~Lu, X.~Xu, Y.~Liu, I.~Weber, L.~Zhu, and W.~Zhang, ``ubaas: A unified
  blockchain as a service platform,'' \emph{Future Generation Computer
  Systems}, vol. 101, pp. 564--575, 2019.

\bibitem{decision_model}
X.~Xu, H.~M. N.~D. Bandara, Q.~Lu, I.~Weber, L.~Bass, and L.~Zhu, ``A decision
  model for choosing patterns in blockchain-based applications,'' in \emph{18th
  IEEE Int. Conf. on Software Architecture (ICSA 2021)}, 2021.

\bibitem{smartmoney}
\BIBentryALTinterwordspacing
D.~Royal, N.~Lim, M.~Staples, P.~Rimba, and S.~Gilder, ``Making money smart:
  Indicative data analytics that could be supported by the blockchain proof of
  concept,'' 2018. [Online]. Available:
  \url{https://data61.csiro.au/en/Our-Research/Our-Work/SmartMoney}
\BIBentrySTDinterwordspacing

\bibitem{smartmoneyfinland}
\BIBentryALTinterwordspacing
TietoEvry and Kela, ``Smart money specification document v1.0,'' 2019.
  [Online]. Available:
  \url{https://www.tietoevry.com/en/campaigns/2020/smartmoney--a-conditional-digital-payment-guarantee/#Smart-money-51723}
\BIBentrySTDinterwordspacing

\bibitem{escrow}
S.~Goldfeder, J.~Bonneau, R.~Gennaro, and A.~Narayanan, ``Escrow protocols for
  cryptocurrencies: {H}ow to buy physical goods using {B}itcoin,'' in
  \emph{Financial Cryptography and Data Security}, A.~Kiayias, Ed.\hskip 1em
  plus 0.5em minus 0.4em\relax Cham: Springer International Publishing, 2017,
  pp. 321--339.

\bibitem{meszaros1998pattern}
D.~J. Meszaros and J.~Doble, ``A pattern language for pattern writing,'' in
  \emph{Proc. Intl. Conf. on Pattern Languages of Program Design}, Oct. 1997.

\bibitem{lorikeet2020}
Q.~Lu \emph{et~al.}, ``Integrated model-driven engineering of blockchain
  applications for business processes and asset management,'' \emph{Software:
  Practice and Experience}, pp. 1--21, 2020.

\bibitem{Perth_Mint}
\BIBentryALTinterwordspacing
``{Perth Mint} gold token,'' {Trovio Operating Pty Ltd.}, Tech. Rep., Mar.
  2021. [Online]. Available:
  \url{https://pmgt.io/static/assets/pmgt_whitepaper.pdf}
\BIBentrySTDinterwordspacing

\bibitem{carpolicy2018}
L.~Bader, J.~C. Bürger, R.~Matzutt, and K.~Wehrle, ``Smart contract-based car
  insurance policies,'' in \emph{2018 IEEE Globecom Workshops (GC Wkshps)},
  2018, pp. 1--7.

\bibitem{Merchant_Token}
\BIBentryALTinterwordspacing
J.~Cavebring, A.~Arvidsson, and T.~Buenvenida, ``{DeFi} merchant payment
  protocol with consumer protection for web 3.0,'' Tech. Rep., Mar. 2021.
  [Online]. Available:
  \url{https://static.hips.com/pdf/white_paper/merchant_token_white_paper.pdf}
\BIBentrySTDinterwordspacing

\bibitem{Lightning_Network}
\BIBentryALTinterwordspacing
J.~Poon and T.~Dryja, ``The {Bitcoin Lightning Network}: {S}calable off-chain
  instant payments,'' Tech. Rep., Jan. 2016. [Online]. Available:
  \url{http://lightning.network/lightning-network-paper.pdf}
\BIBentrySTDinterwordspacing

\end{thebibliography}
